\documentclass[11pt,a4paper]{article}
\usepackage[utf8]{inputenc}
\usepackage{a4wide}
\usepackage{amsmath}
\usepackage{amssymb}
\usepackage{amsfonts}
\usepackage{cite}
\usepackage{color}
\usepackage{adjustbox}
\usepackage{multirow}
\usepackage{pdflscape}
\usepackage{textcomp}
\usepackage{gensymb}
\usepackage{graphicx}
\usepackage{diagbox}
\usepackage{pifont}
\usepackage{bigstrut}
\usepackage[small,bf]{caption}
\usepackage{tabulary}
\usepackage{booktabs}
\usepackage{stackengine}
\usepackage{mathtools}
\usepackage{enumerate}
\usepackage{makecell}
\usepackage{listings}
\usepackage{bbm}
\usepackage[unicode]{hyperref}

\definecolor{greenLinks}{rgb}{0, 0.6, 0} 
\definecolor{blueLinks}{rgb}{0, 0, 0.6}
\definecolor{redLinks}{rgb}{0.6, 0, 0}
\definecolor{eprintLinks}{rgb}{0.4, 0.4, 0.4}
\definecolor{journalLinks}{rgb}{0.6, 0, 0}
\newcommand{\MYhref}[3][redLinks]{\href{#2}{\color{#1}{#3}}}%

\newcommand{\fig}[1]{fig.~\ref{fig:#1}}

\newcommand{\tab}[1]{table~\ref{tab:#1}}
\newcommand{\Eq}[1]{Eq.~(\ref{eq:#1})}

\allowdisplaybreaks
\numberwithin{equation}{section}
\makeatletter
\g@addto@macro\bfseries{\boldmath}
\makeatother

\begin{document}

\begin{titlepage}

 \vspace*{8mm}

\begin{center}
{ \bf\LARGE { Natural inflation with a nonminimal coupling to gravity}}\\[8mm]
Yakefu Reyimuaji $^{\,a, }$\footnote{ yakefu@mail.itp.ac.cn}, 
Xinyi Zhang $^{\,b, c, }$\footnote{corresponding author: zhangxinyi@ihep.ac.cn }\\ 

\vspace{8mm}
$^{a}$\,{\it CAS Key Laboratory of Theoretical Physics, Institute of Theoretical Physics, Chinese Academy of Sciences, Beijing 100190, China} \\
\vspace{2mm}
$^{b}$\,{\it School of Physics and State Key Laboratory of Nuclear Physics and Technology,\\ Peking University, Beijing 100871, China} \\
\vspace{2mm}

$^{c}$\,{\it Institute of High Energy Physics, Chinese Academy of Sciences, Beijing 100049, China} \\
\vspace{2mm}

\end{center}
\vspace{8mm}

\begin{abstract}
\noindent Although natural inflation is a theoretically well-motivated model for cosmic inflation, it is in tension with recent Planck cosmic microwave background anisotropy measurements. We present a way to alleviate this tension by considering a very weak nonminimal coupling of the inflaton field to gravity in both contexts of metric and Palatini formulations of general relativity. We start our discussions with a generic form of the inflaton coupling to the Ricci scalar, then focus on a simple form to do phenomenological study. Our results show that such an extension can bring natural inflation's predictions to a good agreement with the Planck data. Depending on values of the coupling constant $\xi$ and the symmetry breaking scale $f$, we find that with $|\xi|\sim 10^{-3}$ and $f\gtrsim 2.0 M_{\mathrm{pl}}$ predictions of the model stay inside $68\%$ CL allowed region until  $f$ increases up to $7.7 M_{\mathrm{pl}}$, then only inside $95\%$ CL region after $f$ exceeds the latter value. The predictions from the metric and the Palatini theories are very similar due to the simple form of the coupling function we use and the small magnitude of the coupling $\xi$. Successful reheating can also be realized in this model.

\end{abstract}

\end{titlepage}
\setcounter{footnote}{0}

\section{Introduction}
\label{sec:intro}

Inflation in the standard cosmology, as the name suggests, is a theory of an accelerated expansion of the Universe at its very early stage~\cite{Guth:1980zm,Linde:1981mu,Albrecht:1982wi,Linde:1983gd}. From the particle physics point of view, the driving force of such an expansion can be a slowly rolling  scalar field (or fields), called inflaton, moving towards the minimum of its potential. This elegant mechanism provides solutions to several problems that the big-bang cosmology faces with, generates initial seeds needed for late-time structure formation through quantum fluctuations of the scalar field around the classical background, and, in the end, leads the Universe into a radiation-dominated era by a process called reheating~\cite{Kofman:1997yn}. Based on the idea, there are many models of inflation (see ~\cite{Brandenberger:1984cz,Lidsey:1995np,Lyth:1998xn,Bassett:2005xm,Martin:2013tda} for reviews), among which the simplest models rely on a single inflaton field $\phi$ and on the assumption that there is no interaction of the inflaton with other fields except for a minimal coupling to the background metric,
\begin{equation}\label{eq:simpact}
S=\int \mathrm{d}^{4} x \sqrt{-g}\left[\frac{M_{\mathrm{pl}}^2}{2} R+\frac{1}{2} g^{\mu \nu} \partial_{\mu} \phi \partial_{\nu} \phi-V(\phi)\right].
\end{equation}
Physics of these inflationary models is based on the action which is a sum of the Einstein-Hilbert action and an action describing the inflaton field's self interaction. Identification of each model relies on a specification of the potential. Although many theoretically well-motivated models have been proposed so far, some of them are already excluded by disagreement between their predictions, in particular, on scalar spectral index $n_s$ and tensor-to-scalar ratio $r$ and results from the cosmological observations; for instance, the Planck Cosmic Microwave Background radiation (CMB) anisotropy measurements~\cite{Akrami:2018odb}. This means they might not be successful candidates in their simplest forms, constructed by presuppositions of having a canonical kinetic term and a minimal coupling to the metric as in \Eq{simpact}, but they might work well in slightly modified scenarios, such as nonminimal coupling to gravity~\cite{Fakir:1990eg,Faraoni:1996rf,Linde:2011nh,Salvio:2017xul,Almeida:2018pir,Ferreira:2018nav,Antoniadis:2018yfq,Salvio:2019wcp,Simeon:2020lkd,Takahashi:2020car}, introducing multiple inflaton fields~\cite{Linde:1993cn,Copeland:1994vg,Liddle:1998jc,Wands:2007bd} or allowing interactions with other fields~\cite{Berera:1995wh,Berera:1995ie,Mohanty:2008ab,Visinelli:2011jy,Mishra:2011vh,Reyimuaji:2020bkm}. In the present work we study the former possibility with a focus on the natural inflation model.

Natural inflation~\cite{Freese:1990rb,Adams:1992bn} is a model characterized by the axion-like potential
\begin{equation}\label{eq:simpppot}
 V(\phi)=\Lambda^{4}\left(1+\cos \frac{\phi}{f}\right),
\end{equation}
where $\Lambda$ is a scale of an effective field theory generating this potential, and $f$ is a symmetry breaking scale. It is an attractive model due to some nice properties: the inflaton field naturally arises in particle physics as a pseudo Nambu-Goldstone boson from a spontaneously broken global symmetry. Besides, the inflaton has an axion-like origin. Thus it possesses a shift symmetry, which protects the potential from getting large radiative corrections. Despite having such intriguing features, this model is disfavored by the Planck CMB observations~\cite{Akrami:2018odb}. Although changing assumptions about neutrino properties in the analysis of Planck 2015 data shifts slightly the allowed contour towards the prediction of the model, the improvement is lost once the measurements of the B-mode power spectrum and baryon acoustic oscillation are included~\cite{Gerbino:2016sgw}, and the tension is still there. So it is timely to study the model in a slightly different context, particularly, a nonminimal coupling to gravity.

Natural inflation with a periodic nonminimal coupling has been studied in~\cite{Ferreira:2018nav,Simeon:2020lkd}, which are based on the assumptions that the coupling function to gravity possesses a periodicity of $2\pi f$ in field $\phi$ as the potential in \Eq{simpppot} does. This coupling function vanishes at a minimum of the potential. However, in our opinion, these requirements can be relaxed because there is no symmetry protecting the periodicity. Besides, an extended theory with a nonminimal coupling to gravity can realize the reheating either relying on the usual reheating mechanism in which the inflaton field oscillates and decays at the minimum of the potential or, instead, with gravitational particle production~\cite{Parker:1969au,Ford:1986sy} if the minimum of the potential gets modified or even without the minimum (see two such examples in~\cite{Takahashi:2020car}). In this work, we start with a generic coupling to the Ricci scalar alone and study its consequences on the predictions of the cosmological observables by using its simplest form in both metric and Palatini theories of gravity. 

The outline of the paper is the following. In section~\ref{overview} there is an overview about the nonminimal coupling of natural inflaton in the context of the scalar-tensor theory of gravity. Section~\ref{sec:metcase} and~\ref{sec:palcase} encompass results in metric and Palatini formulations of gravity, respectively. Further discussions about the results and reheating after the inflation are given in section \ref{sec:discussions}. Section \ref{sec:conclusions} draws conclusions of the paper.

\section{Nonminimal coupling of inflaton to gravity}
\label{overview}
As a next-to-simplest set-up, we consider an extension of the action~\Eq{simpact} to 
\begin{equation}\label{eq:nextsimpact}
S=\int \mathrm{d}^{4} x \sqrt{-g}\left[\frac{M_{\mathrm{pl}}^2}{2}F(\xi,\phi) R+\frac{1}{2} g^{\mu \nu} \partial_{\mu} \phi \partial_{\nu} \phi-V(\phi)\right],
\end{equation}
where we introduce a nonminimal coupling function $F(\xi,\phi)$, but keep other parts of the action in \Eq{simpact} unchanged. Here, $\xi$ is the coupling between the inflaton $\phi$ and the Ricci scalar $R$, $g$ is the determinant of the spacetime metric $g_{\mu\nu}$, and $M_{\mathrm{pl}}=\frac{1}{\sqrt{8\pi G}}$ is the reduced Planck mass. This action describes a type of scalar-tensor theories of gravity formulated in the Jordan conformal frame, which, in general, is referred to as a generalization of the Einstein-Hilbert action by introducing a scalar field with a nonminimal coupling to gravity, like in \Eq{nextsimpact}. It may also include a non-canonical kinetic term. In our discussions, we keep the kinetic term in canonical form and leave the function $F(\xi,\phi)$ to be determined.  

One can also write the action in the Einstein frame of the scalar-tensor theory, which is related to the Jordan frame by the Weyl (or conformal) transformation of the metric. In this frame, the nonminimal coupling to gravity is implicit. These two frames can be regarded as mathematically dual descriptions of the same theory. In what follows, we choose to work in the Einstein frame because it is formally analogous to the minimal coupling case, and computations are relatively straightforward.

There are two different formulations of general relativity. One is the metric theory~\cite{Sebastiani:2016ras}, where the metric $g_{\mu\nu}$ is an independent variable and all the other quantities like the connection and the curvature tensor are obtained by the metric and its derivatives. The other one is the Palatini theory~\cite{Tenkanen:2020dge}, in which the spacetime metric and the connection are considered as independent variables. These two formulations lead to different predictions on the cosmological observables~\cite{Bauer:2008zj,Bauer:2010jg,Tamanini:2010uq,Gialamas:2019nly}. The scalar field arising from the Weyl conformal transformation has different dependencies on the nonminimal coupling function in these two descriptions, which will be clear later in this work. 

To proceed, inflation in the Einstein frame is related to the Jordan frame by the Weyl transformation of the metric and redefinition of the scalar field~\cite{Maeda:1988ab,Tenkanen:2020dge},
\begin{equation}
\begin{aligned}\label{eq:conftran}
 \tilde{g}_{\mu\nu} = & F(\xi,\phi) g_{\mu\nu}, \\
  \frac{d \chi}{d \phi}= & \sqrt{\frac{1}{F(\xi, \phi)^2} \left[F(\xi, \phi) +\frac{3k}{2}  M_{\mathrm{pl}}^2 \left( \frac{\partial F(\xi, \phi)}{\partial \phi} \right)^2  \right]} \\
                       = & \sqrt{\frac{1}{F(\xi, \phi)} +\frac{3k}{2}  M_{\mathrm{pl}}^2 \left( \frac{\partial \ln F(\xi, \phi)}{\partial \phi} \right)^2 },
\end{aligned}
\end{equation}
where $\tilde{g}_{\mu\nu}$ is the rescaled metric and $\chi (\phi)$ is the inflaton field in the Einstein frame, $k$ is either 0 or 1 corresponding to the Palatini or metric theory of gravity. By this transformation, the action \eqref{eq:nextsimpact} becomes 
\begin{equation}\label{eq:simpactEinst}
S_\mathrm{E}=\int \mathrm{d}^{4} x \sqrt{-\tilde{g}}\left[\frac{M_{\mathrm{pl}}^2}{2} \tilde{R}+\frac{1}{2} \tilde{g}^{\mu \nu} \partial_{\mu} \chi \partial_{\nu} \chi-V_\mathrm{E}(\chi)\right],
\end{equation}
where
\begin{equation}
 V_\mathrm{E} \left[\chi(\phi)\right] = \frac{V(\phi)}{F(\xi,\phi)^2}.
\end{equation}
As the conformal transformation becomes singular when $F(\xi,\phi)$ vanishes, it should be restricted to the field space where this function has a definite sign. In \Eq{conftran} we take a positive sign for the function $F(\xi, \phi)$, and henceforth stick to this convention. \Eq{conftran} shows that $d\chi/d\phi >0$ for any positive-valued function $F(\xi,\phi)$, so, one can conclude that $\chi$ is a monotonically increasing function for the variable $\phi$, without knowing its expression. Moreover, there is not only one-to-one correspondence between $\chi$ and $\phi$, but also a full space of $\phi$ covers that of $\chi$. This allows us to use $\phi$ as a parameter when we do not know expressions of some quantities in terms of $\chi$. 

Dynamics of the inflaton field and the Friedmann-Robertson-Walker (FRW) geometry is described by
\begin{equation}
 \begin{aligned}
  & \ddot{\chi}+3 H \dot{\chi}+V^\prime_{\rm E}(\chi)=0 , \\
  & \frac{1}{3 M_{\mathrm{pl}}^{2}}\left[\frac{1}{2}\dot{\chi}^{2}+V_{\rm E}(\chi)\right] = H^{2} ,
 \end{aligned}
\end{equation}
where dot and double dots are derivatives with respect to cosmic time $t$, prime is derivative with respect to field $\chi$, $M_{\mathrm{pl}} = \frac{1}{\sqrt{8\pi G }}$ is the reduced Planck mass, and $H$ is the Hubble expansion rate. Inflation with a nonminimal coupling in the Jordan frame, in \Eq{nextsimpact}, can be easily studied by the transformation to the Einstein frame, in which slow-roll parameters are expressed in a similar way as in the case of a minimal coupling,
\begin{equation}
\begin{aligned}
 \epsilon_\mathrm{v} (\chi) = & \frac{M_{\mathrm{pl}}^2}{2}\left[ \frac{V'_\mathrm{E}(\chi)}{V_\mathrm{E}(\chi)} \right]^2, \\
  \eta_\mathrm{v} (\chi) = & M_{\mathrm{pl}}^2 \frac{V''_\mathrm{E}(\chi)}{V_\mathrm{E}(\chi)},
\end{aligned}
\end{equation}
where $V'_\mathrm{E}$ and $V''_\mathrm{E}$ are the first and the second order derivatives of the Einstein frame potential with respect to the field $\chi$. 

Applying above general discussions to the natural inflation, the Einstein frame potential reads
\begin{equation}\label{eq:einstpotent}
  V_\mathrm{E} \left[\chi(\phi)\right] = \frac{\Lambda^{4}}{F(\xi,\phi)^2} \left(1+\cos \frac{\phi}{f}\right).
\end{equation}
In order to have a more flat potential $V_\mathrm{E}$ than the one in the Jordan frame, the inequality $V'_\mathrm{E}(\chi)\le \frac{d V(\phi)}{d \phi}$ has to be satisfied. Before that, $V'_\mathrm{E}(\chi)\le 0$ is required to have a decreasing potential, which translates into $\frac{\partial}{\partial \phi } \frac{V}{F^2} \le 0$, at each point in the field space, for any $F(\xi,\phi)$ that is a finite, positive, and real analytic function. 

Knowing the form of the potential in \Eq{einstpotent}, corresponding slow-roll parameters read
\begin{equation}
\begin{aligned}\label{eq:epsil}
 \epsilon_\mathrm{v} = & \frac{M_{\mathrm{pl}}^2}{2} \left( \frac{d \chi}{d \phi} \right)^{-2} \left(\frac{\partial \ln V_\mathrm{E}}{\partial \phi} \right)^{2} \\
 = & \frac{M_{\mathrm{pl}}^2}{2} \left( \frac{d \chi}{d \phi} \right)^{-2} \left[ 2 \frac{\partial \ln F}{\partial \phi} + \frac{\sin \frac{\phi}{f}}{f\left( 1+\cos \frac{\phi}{f}  \right)}   \right]^2, 
\end{aligned}
\end{equation}
which is non-negative, and
\begin{equation}\label{eq:etttta}
\begin{aligned}
\eta_\mathrm{v}= &  M_{\mathrm{pl}}^2 \left( \frac{d \chi}{d \phi} \right)^{-2} \left\{ \frac{3}{2}  \frac{\partial \ln F}{\partial \phi} \left[ 1+ \frac{k}{2} M_{\mathrm{pl}}^2 \left( \frac{d \chi}{d \phi} \right)^{-2} \left[ \left( \frac{\partial \ln F}{\partial \phi} \right)^2 + 2 \frac{\partial^2 \ln F}{\partial \phi^2}\right]   \right]  \right. \\ 
& \times \left[ 2 \frac{\partial \ln F}{\partial \phi} + \frac{\sin \frac{\phi}{f}}{f\left( 1+\cos \frac{\phi}{f}  \right)}   \right] +  2 \frac{\partial \ln F}{\partial \phi} \frac{\sin \frac{\phi}{f}}{f\left( 1+\cos \frac{\phi}{f}  \right)} \\
&\left. -  2 \frac{\partial^2 \ln F}{\partial \phi^2} - \frac{\cos \frac{\phi}{f}}{f^2\left( 1+\cos \frac{\phi}{f}  \right)} \right\}.
\end{aligned}
\end{equation}
Here we omit variables of $F(\xi,\phi)$. We can see from these expressions that if $\xi =0$, then $F =1$ and $ \frac{d \chi}{d \phi} =1$, the expressions of the slow-roll parameters returns to the minimally coupled results
\begin{equation}
 \begin{aligned}
 \epsilon_\mathrm{m, v} = & \frac{M_{\mathrm{pl}}^2}{2}  \frac{\sin^2 \frac{\phi}{f}}{f^2\left( 1+\cos \frac{\phi}{f}  \right)^2}, \\
 \eta_\mathrm{m, v}= & - M_{\mathrm{pl}}^2 \frac{\cos \frac{\phi}{f}}{f^2\left( 1+\cos \frac{\phi}{f}  \right)},
 \end{aligned}
\end{equation}
where the first subscript stands for a minimal coupling. 

The number of e-folds is given by 
\begin{equation}\label{eq:Nefold}
 N = \int^{\chi_{\text {end }}}_{\chi_{*}} \frac{H}{ \dot{\chi}} \mathrm{~d} \chi \approx \frac{1}{M_{\mathrm{pl}}^2} \int_{\chi_{\text {end }}}^{\chi_{*}} \frac{ V_\mathrm{E}}{ V'_\mathrm{E}} \mathrm{~d} \chi ,
\end{equation}
where
\begin{equation}
 \begin{aligned}
  H^2 \simeq \frac{ V_\mathrm{E}(\chi)}{3M_{\mathrm{pl}}^2}, \\
  \dot{\chi} \simeq -\frac{ V'_\mathrm{E}(\chi)}{3H},
 \end{aligned}
\end{equation}
are implemented. So far, there is no need to know an analytic expression of the $\chi(\phi)$, its derivative $\frac{d \chi}{d \phi}$ is enough. But one may wonder that performing the integral in \Eq{Nefold} requires to write down the potential $V_\mathrm{E}$ in terms of $\chi$ thus the inverse function $\phi(\chi)$ is needed. We have argued that $\chi(\phi)$ is strictly monotonic function, so its inverse must exist and is also monotonic. This enables us to use $\phi$ as a parameter and perform an integration with it. In case it is not possible to perform the integration in \Eq{Nefold} for reasons that the analytic expression of $\chi(\phi)$ or the inverse function $\phi(\chi)$ cannot be found, an analytic expression of $N$ cannot be written down. But we can still perform
\begin{equation}
  N = \frac{1}{M_{\mathrm{pl}}^2} \int^{\phi_{\text {end }}}_{\phi_{*}} \frac{\frac{1}{F } +\frac{3k}{2}  M_{\mathrm{pl}}^2 \left( \frac{\partial \ln F }{\partial \phi} \right)^2}{2 \frac{\partial \ln F}{\partial \phi} + \frac{\sin \frac{\phi}{f}}{f\left( 1+\cos \frac{\phi}{f}  \right)} }   \mathrm{~d} \phi ,
\end{equation}
and construct a relation between $N$ and $\chi_{*}$ through the parameter $\phi_{*}$.  

The scalar spectral index and the tensor-to-scalar ratio have the following expressions
\begin{equation}\label{eq:cosobsevb}
\begin{aligned}
  n_\mathrm{s} & =  1-6\epsilon_\mathrm{v} + 2\eta_\mathrm{v}, \\
  r & =  16 \epsilon_\mathrm{v},
 \end{aligned}
\end{equation}
in which it is understood that $\epsilon_\mathrm{v}$ and $\eta_\mathrm{v}$ are evaluated at the field value of horizon crossing.

In what follows, we construct a specific model and study its phenomenology. As long as we know an explicit form of the function  $F(\xi,\phi)$, we build up a model of inflation. Based on the model, we can compute the field value at the horizon crossing, and get predictions for the cosmological observables $n_\mathrm{s}$ and $r$, by plugging \Eq{epsil} and \Eq{etttta} into \Eq{cosobsevb}. Hence, an important task is to answer the question which form of $F(\xi,\phi)$ is an appropriate choice. In order to find a specific form the function $F(\xi,\phi)$, we must bear in mind that the inflaton coupling to metric returns back to the minimal coupling scenario when $\xi =0$ irrespective of the field value $\phi$. So, the coupling function $F(\xi,\phi)$ may take forms like $1+\xi f(\phi)$, $a^{\xi f(\phi)}$, $\cos \left[ \xi f(\phi)\right]$, with an unspecified function $f(\phi)$ and a positive constant $a$. On top of these, we consider a small deviation from the minimal coupling scenario, which means the coupling cannot be large, say  $\left| \xi \right| < 1$. Having settled the framework, one may notice that the last two functions above can be expanded around $\xi =0$, respectively, as power series $1+\xi \ln a f(\phi)+ \dots$ and  $1- \frac{1}{2}\xi^2 f(\phi)^2+ \dots$. Therefore, all these three forms can be represented by
\begin{equation}
 F(\xi,\phi) = 1+ \xi f(\phi),
\end{equation}
at the leading order, upon some redefinitions of $\xi$ or $f(\phi)$. To fix the form of the arbitrary function $f(\phi)$, we rely on the following guiding rules. First of all, we suppose that the action preserves CP symmetry so $f(\phi)$ is a function of $\phi^2$. This is due to the fact that $\phi$, as a pseudoscalar, is odd under the parity transformation. Moreover, a necessary condition $f(\phi)$ has to satisfy is $\frac{\partial}{\partial \phi} \left(\frac{V}{\left[ 1+ \xi f(\phi) \right]^2} \right) \le 0$ as discussed earlier. Lastly, for simplicity, we choose to use an elementary function.  Based on these rules, a most favorable form of the function $F(\xi,\phi)$ is 
\begin{equation}\label{eq:Ffunction}
 F(\xi,\phi) = 1+ \xi \left( \frac{\phi}{M_{\mathrm{pl}}} \right)^n,
\end{equation}
where the $n$ takes even numbers. The coupling $\xi$ can be any real number, but it is subject to a constraint $\xi > -\left( \frac{M_{\mathrm{pl}}}{f\pi}  \right)^n$ in order to guarantee that the function $ F(\xi,\phi)$ remains positive in the entire field space $\phi\in [0, f\pi]$. In the following sections we study this nonminimal coupling model, discuss its predictions, and show differences comparing to that in the minimal coupling scenario.

\section{Metric theory}
\label{sec:metcase}
\begin{figure}
  \centering
  \includegraphics[width=0.55\textwidth]{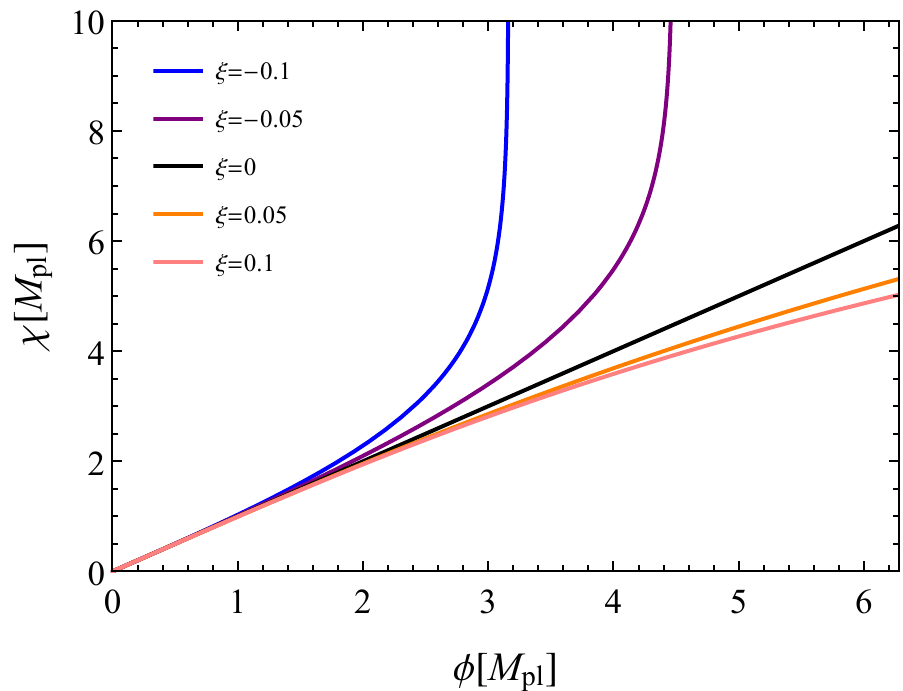}
  \caption{
    Plot of field $\chi$ as a function of $\phi$ with $n=2$. Color lines correspond to different values of the coupling $\xi$.}
  \label{fig:chiphi}
\end{figure}

Specifying the form of $F(\xi, \phi)$ in \Eq{Ffunction}, expression of $\chi$ in terms of a dimensionless parameter $x\equiv \frac{\phi}{M_{\mathrm{pl}}}$ is
\begin{equation}
 \sqrt{\xi} \frac{\chi}{M_{\mathrm{pl}}}  = \sqrt{1+6\xi} \sinh^{-1} \left( x \sqrt{(1+6\xi)\xi} \, \right)  -\sqrt{6\xi} \tanh^{-1} \left( \frac{\sqrt{6} \xi  x}{\sqrt{1+ (1+6 \xi)\xi  x^2}} \, \right) ,
\end{equation}
for $n=2$, and no analytic expressions can be found for higher values of $n$. The relation between the $\chi$ and $\phi$ is illustrated in \fig{chiphi}. As shown, $\chi$ is an increasing function of $\phi$ for all values of the coupling constant $\xi$, but it shoots up with a negative coupling compared to the mild increase with positive values. This is what indicated in \Eq{conftran}, because derivative term contribution is the same  for both positive and negative couplings, but the first term leads to distinction $\frac{d\chi}{d\phi}\ge 1$ for $\xi\le 0$ and $0<\frac{d\chi}{d\phi}\le 1$ for $0<\xi\le \frac{1}{6}$, since we choose Weyl transformation in $F(\xi, \phi)>0$ branch. Moreover, $\chi$ coincides with $\phi$  as the coupling switches off.
 As yet, we make plots of $\chi$ in $n=2$ case and demonstrate these properties. The similar statement can apply to any $n>2$. Despite having no analytic expressions of $\chi$ for $n>2$, numerical integration is still possible, and we confirm its behavior conforms to the expectation.  
 
\begin{figure}[h]
  \centering
  \includegraphics[width=0.9\textwidth]{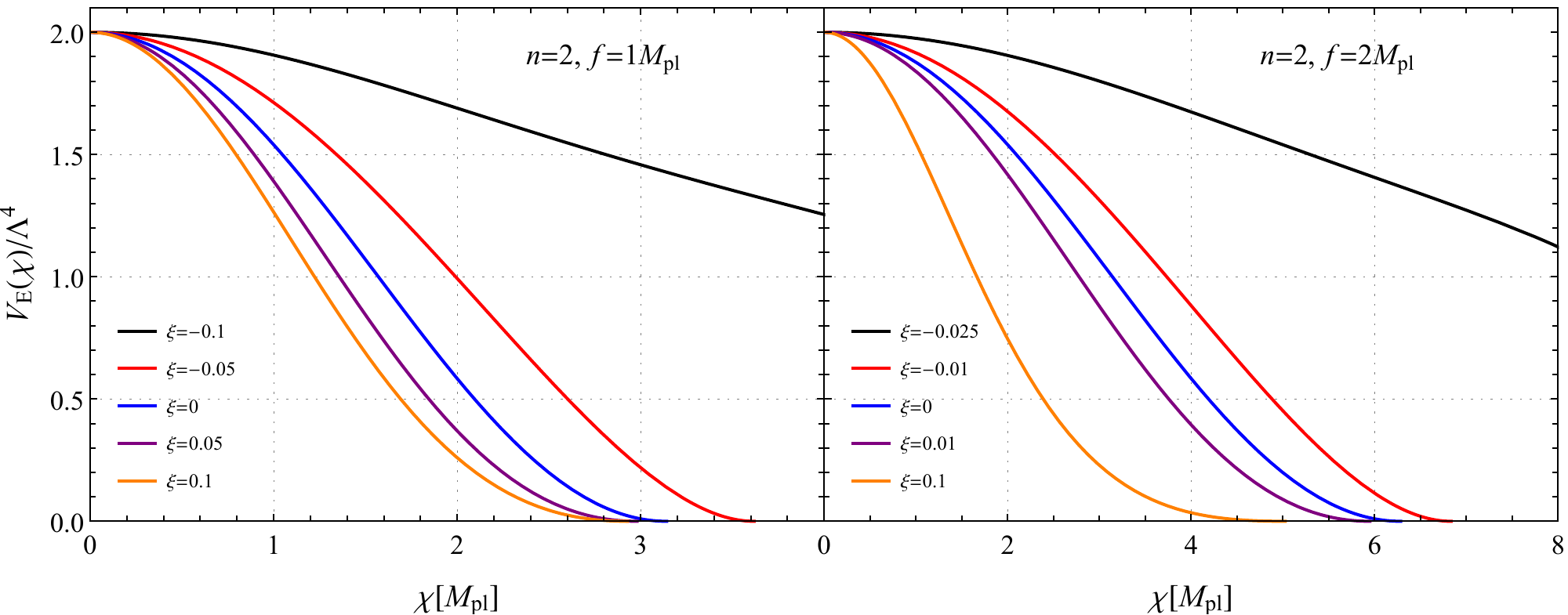}  \\ \vspace{5mm}
      \includegraphics[width=0.9\textwidth]{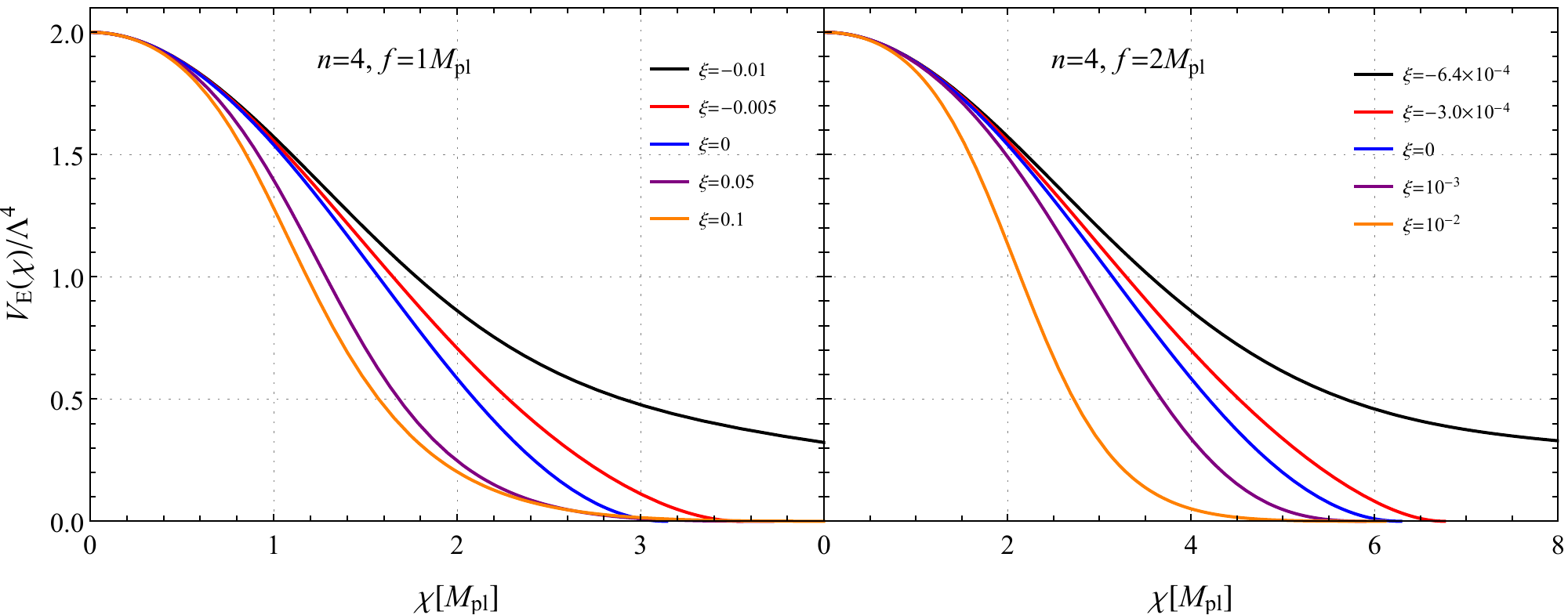}
  \caption{
    Ratio of the Einstein frame potential to fourth power of its scale, $V_\mathrm{E}/\Lambda^4$, as a function of $\chi$ with different benchmark values of the coupling parameter $\xi$. Shown are also the changes with respect to different values of $n$ and $f$.}
  \label{fig:Vechi}
\end{figure}

\begin{figure}[h]
  \centering
  \includegraphics[width=0.48\textwidth]{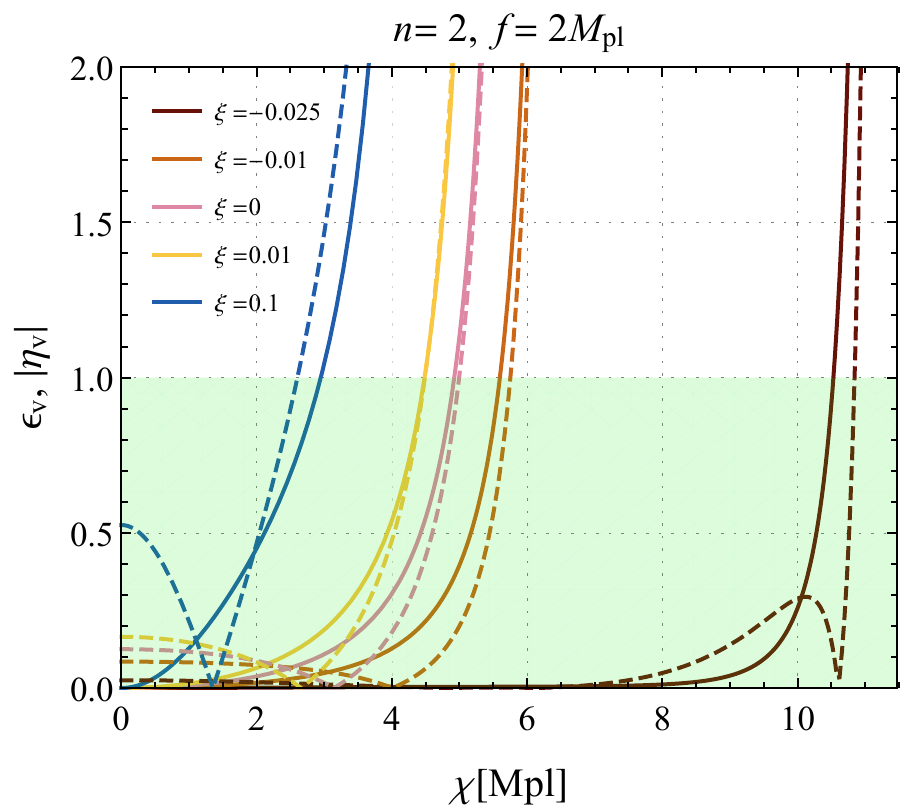} \hspace{2mm}
    \includegraphics[width=0.48\textwidth]{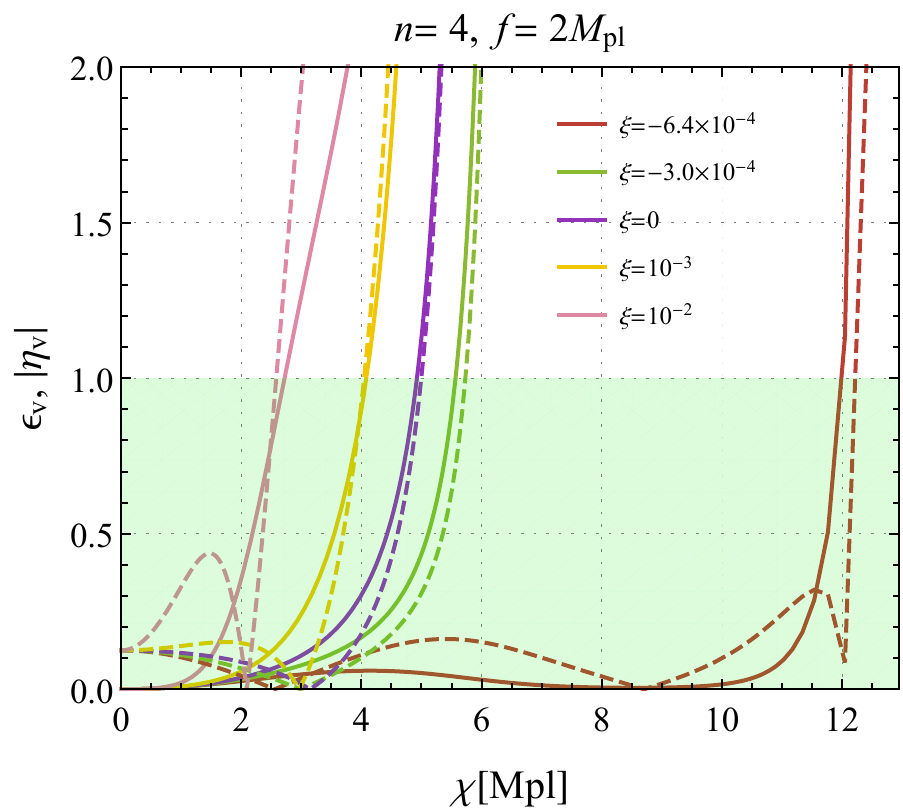}
  \caption{
    The slow-roll parameters $\epsilon_\mathrm{v}$ (solid lines) and $\eta_\mathrm{v}$ (dashed lines) as functions of $\chi$ for the cases of $n=2,4$ and $f=2M_{\mathrm{pl}}$ with some benchmark values of $\xi$. Shaded band is the region where slow-roll conditions are satisfied.}
  \label{fig:slrfigs}
\end{figure}

The Einstein frame potential is given by
\begin{equation}
 V_\mathrm{E} = \Lambda^4 \frac{1+\cos \left(\frac{x}{\hat{f} }\right)}{\left(1+\xi  x^n\right)^2},
\end{equation}
where $\hat{f} \equiv f/M_{\mathrm{pl}}$ is used to avoid carrying dimensionful parameter inside a function. Change in the potential with respect to the inflaton $\chi$ and coupling $\xi$ is illustrated in \fig{Vechi}. The potential gets flatter when the coupling $\xi$ decreases, or the symmetry breaking scale $f$ increases, or the power $n$ decreases, when we consider a change in only one of these variables while keeping others fixed. But, remember that once we fix two of them the third one gets constrained by  the relation $\xi>-\left( \frac{M_{\mathrm{pl}}}{f\pi}  \right)^n$, as argued in the previous section. For instance, although the potential becomes flatter with a smaller negative coupling, we cannot achieve arbitrary flat potential by taking its small value. Because the lower bound of $\xi$ is determined once the $f$ and $n$ are fixed, it prevents $\xi$ from taking values beyond that bound. 
Quantitative measures for flatness of the potential are the slow-roll parameters
\begin{equation}
 \begin{aligned}
  \epsilon_\mathrm{v} = & \frac{\left[ (F-1)\left(2nV/\Lambda^4 + \frac{x}{\hat{f}} \sin \frac{x}{\hat{f}}\right) + \frac{x}{\hat{f}} \sin \frac{x}{\hat{f}} \right]^2 }{(V/ \Lambda^{4})^2 \left[ 2 x^2 F + 3n^2 (F-1)^2\right]} \\ 
                      = & \frac{\left[2  n \xi  x^n+ \left(1+\xi  x^n \right) \frac{x}{\hat{f}} \sin \left(\frac{x}{\hat{f}}\right) +2  n \xi  x^n \cos \left(\frac{x}{\hat{f}}\right)\right]^2}{2 x^2 \left( 1 + \xi  x^{n} + \frac{3}{2} n^2 \xi^2 x^{2 (n-1)} \right)\left[1+ \cos \left(\frac{x}{\hat{f}}\right) \right]^2},
  \end{aligned}
\end{equation}
and 
\begin{equation}
 \begin{aligned}
  \eta_\mathrm{v} = & \frac{ 2( A_1 + A_2) }{(V/ \Lambda^{4}) \left[ 2 x F + 3n^2 (F-1)^2\right]^2},
\end{aligned}
\end{equation}
where
\begin{equation}
 \begin{aligned}
  A_1 = & -\cos \frac{x}{\hat{f} }\left[n x^2 F(F-1)\left( 2 \left[ 2(n-1)-(2+3 n)(F-1) \right] + \frac{3 n}{\hat{f}^{2}} F(F-1)\right)\right. \\
        & \left. + \frac{2 x^4}{\hat{f}^2} F^3 -12n^4(F-1)^4\right], \\
  A_2 = & \ n (F-1)\bigg[ 12  n^3 (F-1)^3 + 2x^2  F  \left(2-2n +(2+ 3 n)(F-1) \right)  \\
        &  \left. + F \frac{x}{\hat{f}}  \sin  \frac{x}{\hat{f}}  (3n (F-1) (n-1 +(4n-1)( F -1) )+7 x^{2} F ) \right].
 \end{aligned}
\end{equation}
In all these expressions $F$ is our chosen function $F(\xi,\phi)=1+\xi x^n$, and $V$ is the potential given in \Eq{simpppot}, their variables are omitted for brevity. Note that in the expressions of the slow-roll parameters above the combination $V/\Lambda^4$ always comes together thus the parameters are independent of the scale $\Lambda$. Having the expressions at hand, it is easy to show the field space where the slow-roll conditions $\epsilon_\mathrm{v}\le 1$ and $\eta_\mathrm{v}\le 1$ are satisfied. For the illustrative purposes, in \fig{slrfigs} we draw the plots of these slow-roll parameters when $f=2M_{\mathrm{pl}}$ and $n=2,4$.
It is apparent from this figure that the field space of $\chi$ satisfying the slow-roll conditions  gets larger (smaller) when $\xi$ decreases (increase). This also confirms the implications in \fig{Vechi}. Now that we have expressions of these slow-roll parameters, the field value at the end of inflation is easily determined by $\epsilon_\mathrm{v}, \eta_\mathrm{v} = 1$, according to the one that satisfies the condition first. As usual, field value at the horizon crossing is computed by equating $N$ to a required number of e-folds.

\begin{figure}[h]
  \centering
  \includegraphics[width=0.49\textwidth]{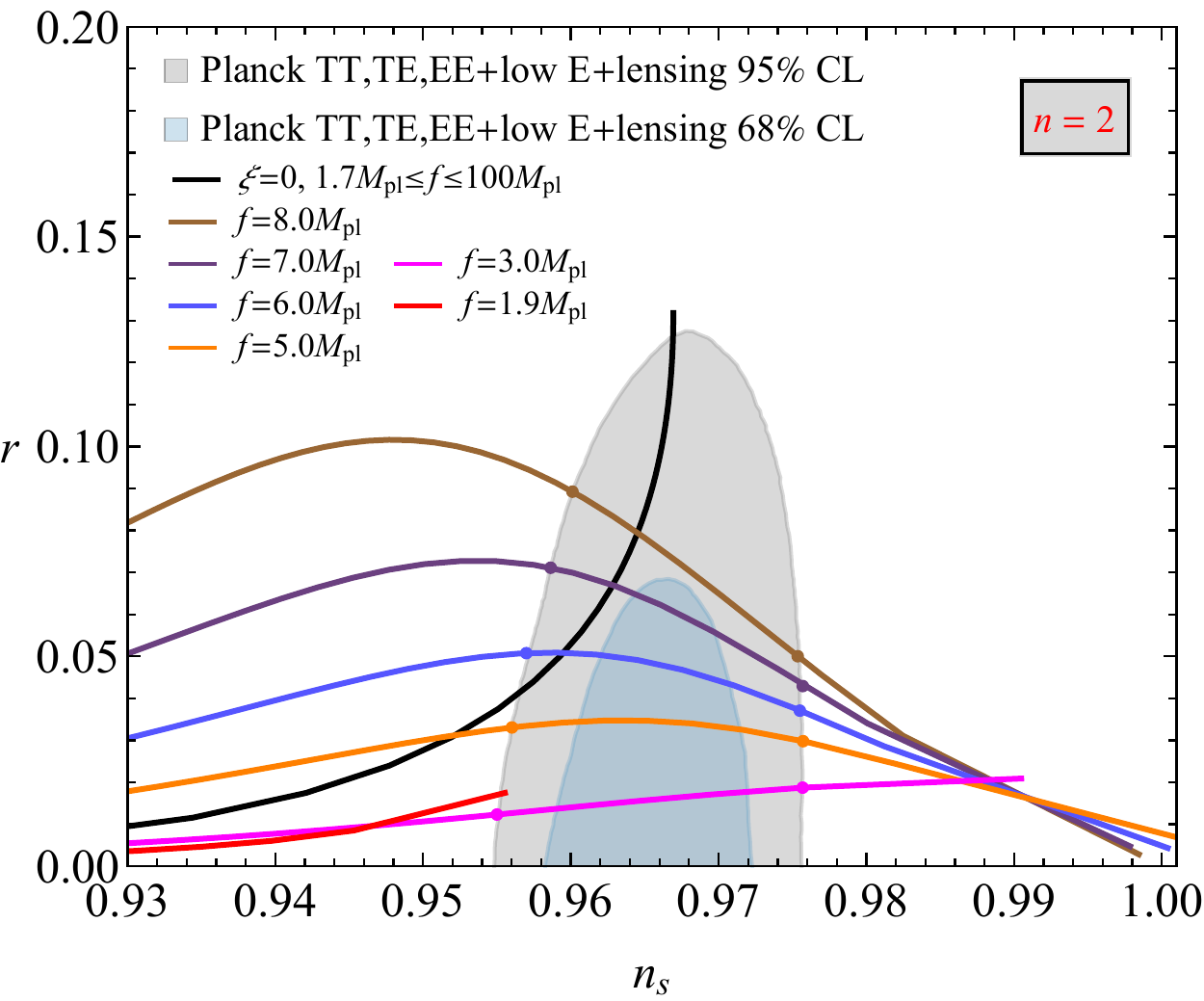}\hspace{1.5mm}
  \includegraphics[width=0.49\textwidth]{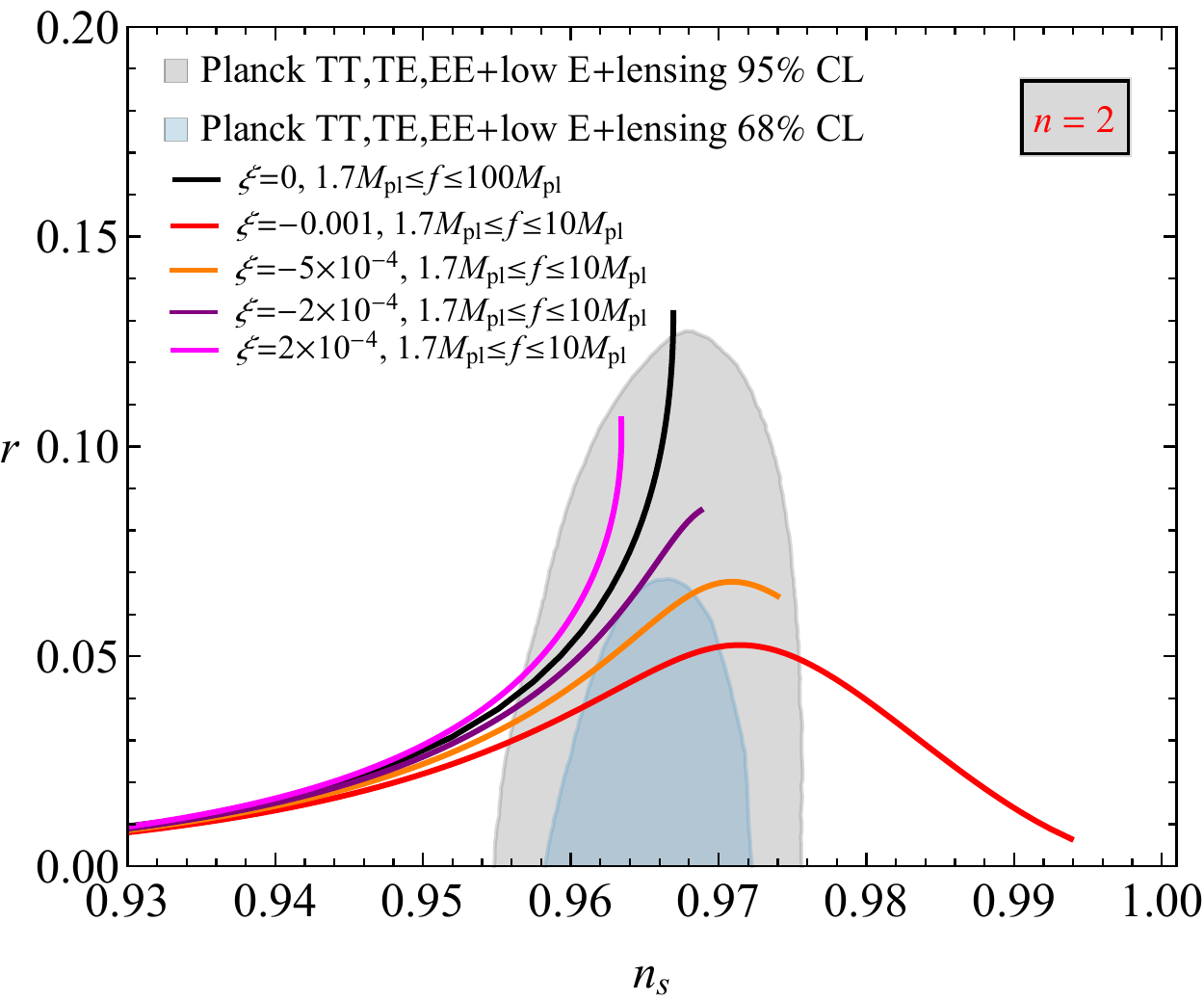} \\ \vspace{3mm}
  \includegraphics[width=0.49\textwidth]{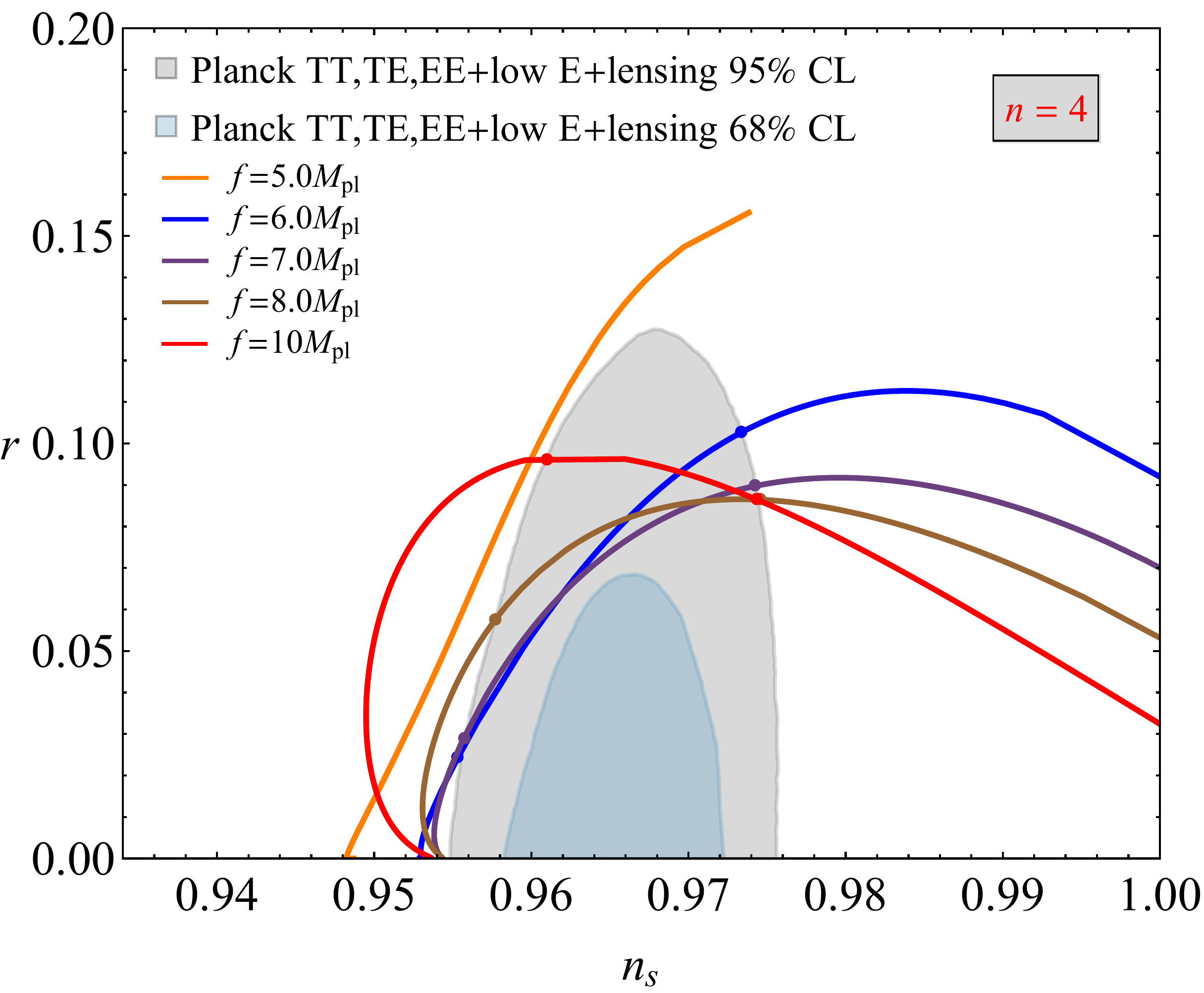}\hspace{1.5mm}
  \includegraphics[width=0.49\textwidth]{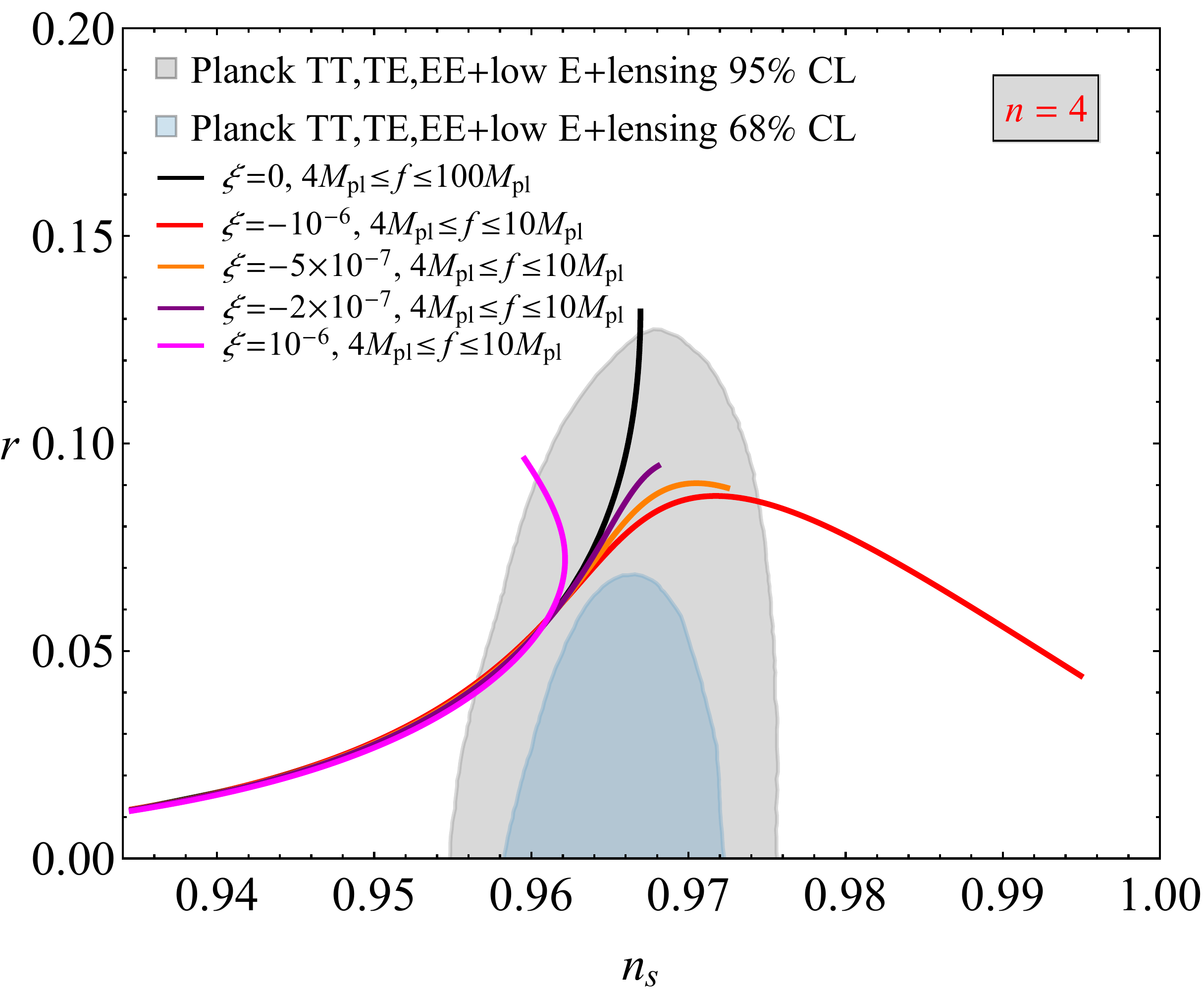}
  \caption{Plots of $n_s$ and $r$ obtained by the natural inflation having a nonminimal coupling to gravity in the metric theory. This figure shows the results when $n=2,4$ and 60 e-folds are achieved before the end of inflation.}
  \label{fig:nsrplot}
\end{figure}

Predictions of this model for the scalar spectral index $n_s$ and the tensor-to-scalar ratio $r$ are shown in \fig{nsrplot}. Each line corresponds to the 60 e-folds that happen before the inflation ends. In the left panels, we illustrate results by fixing $f$ but varying $\xi$. The left edges of the lines indicate the coupling $\xi$ has the largest (or positive) values, while the right ends correspond to the smallest (or negative) values. The black curve corresponds to $\xi =0$. When $n=2$ and $f < 1.9 M_{\mathrm{pl}}$ (the upper-left plot), curves cannot cross over the allowed region of Planck result, for the larger values $2.0 M_{\mathrm{pl}} \le f \le 7.7 M_{\mathrm{pl}}$ they covers both $95\%$ and $68\%$ CL regions, otherwise they just stay in $95\%$ CL contour but not in $68\%$ CL contour. Moreover, the scalar spectral index decreases with the increase of $\xi$.
 When $n=4$ and $f < 5.1 M_{\mathrm{pl}} $ (the down-left plot), there is no chance to have a viable inflation model, whereas the curves obtained from $5.1 M_{\mathrm{pl}} \le f \le 10.0 M_{\mathrm{pl}} $ can enter the $95\%$ CL region but they are still outside the $68\%$ CL region. Right panels are the results for $r$ vs $n_s$ from fixing $\xi$ but varying $f$ in $[1.7,10]M_{\mathrm{pl}}$ for $n=2$ case and in $[4.0,10]M_{\mathrm{pl}}$ for $n=4$ case. As we can see, smaller value of $\xi$ gives a better agreement with the cosmological observations.

\section{Palatini theory}
\label{sec:palcase}
In the Palatini theory of gravity, \Eq{conftran} provides the relation between inflaton fields in the Einstein and the Jordan frames, which can be written as~\cite{Jarv:2017azx}
\begin{equation}
 \frac{\chi}{M_{\mathrm{pl}}} = x \ _2F_1\left(\frac{1}{2},\frac{1}{n};1+\frac{1}{n};-x^n \xi \right),
\end{equation}
where $x\equiv \frac{\phi}{M_{\mathrm{pl}}}$ and the hypergeometric function $_{2}F_{1}(a,b;c;z)$ is defined for arbitrary real numbers $a, b, c$ and $\left|z\right| < 1$ by a series expansion
\begin{equation}
 { }_{2} F_{1}(a, b ; c ; z)=\sum_{k=0}^{\infty} \frac{a_{k}b_{k}}{c_{k}} \frac{z^{k}}{k !},
\end{equation}
 in which the coefficient $\alpha_k$ with $\alpha = a, b ,c$ is computed by $\alpha_k = \displaystyle \prod^{k-1}_{i=0} \left( \alpha +i \right) $ for $k\ge 1$ and $\alpha_0 = 1$ for $k=0$.
 Unlike the case of metric gravity, we can write expression of $\chi$ for any value of $n$. For $n=2$ we can reexpress $\chi$ by
 \begin{equation}
\sqrt{\xi} \frac{\chi}{M_{\mathrm{pl}}}  = \sinh^{-1} \left( x \sqrt{\xi} \right),
 \end{equation}
according to a special case of the hypergeometric function, whereas no such a simplification can be obtained for higher values of $n$. Also, this is an example where we can write the inverse function $\phi(\chi)$ of the inflaton field $\chi$, and express everything just by $\chi$, without resorting to the parametric relation with $\phi$. 

\begin{figure}[h]
  \centering
  \includegraphics[width=0.55\textwidth]{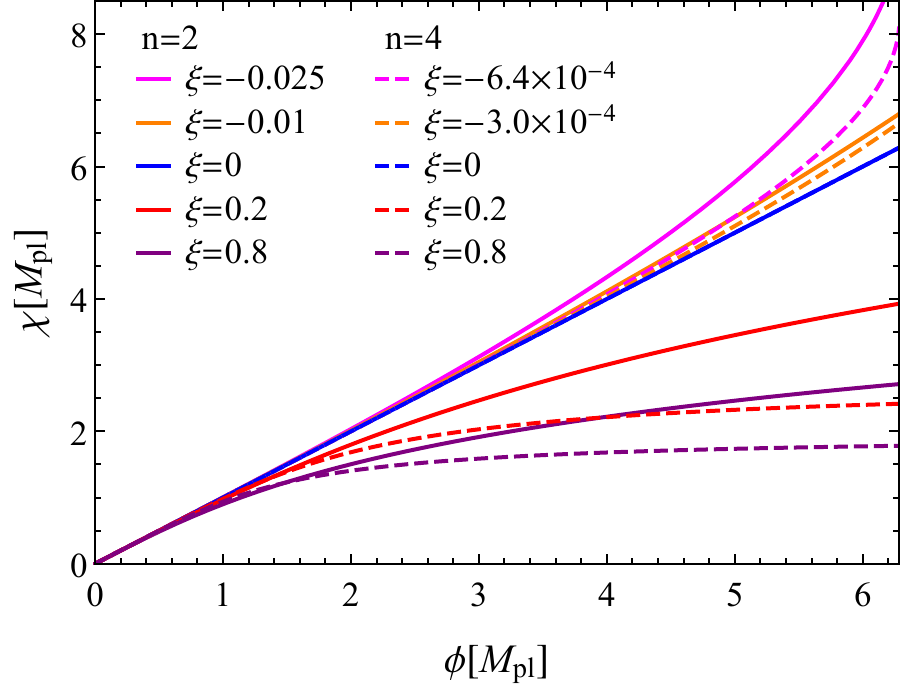}
  \caption{
    Relations between the inflaton fields $\chi$ and $\phi$ with $n=2$ (solid lines) and $n=4$ (dashed lines) in the context of Palatini theory. Each color represents fixed value of $\xi$ indicated inside the plot.}
  \label{fig:chiphipal}
\end{figure}

\begin{figure}[h]
  \centering
  \includegraphics[width= 0.9\textwidth]{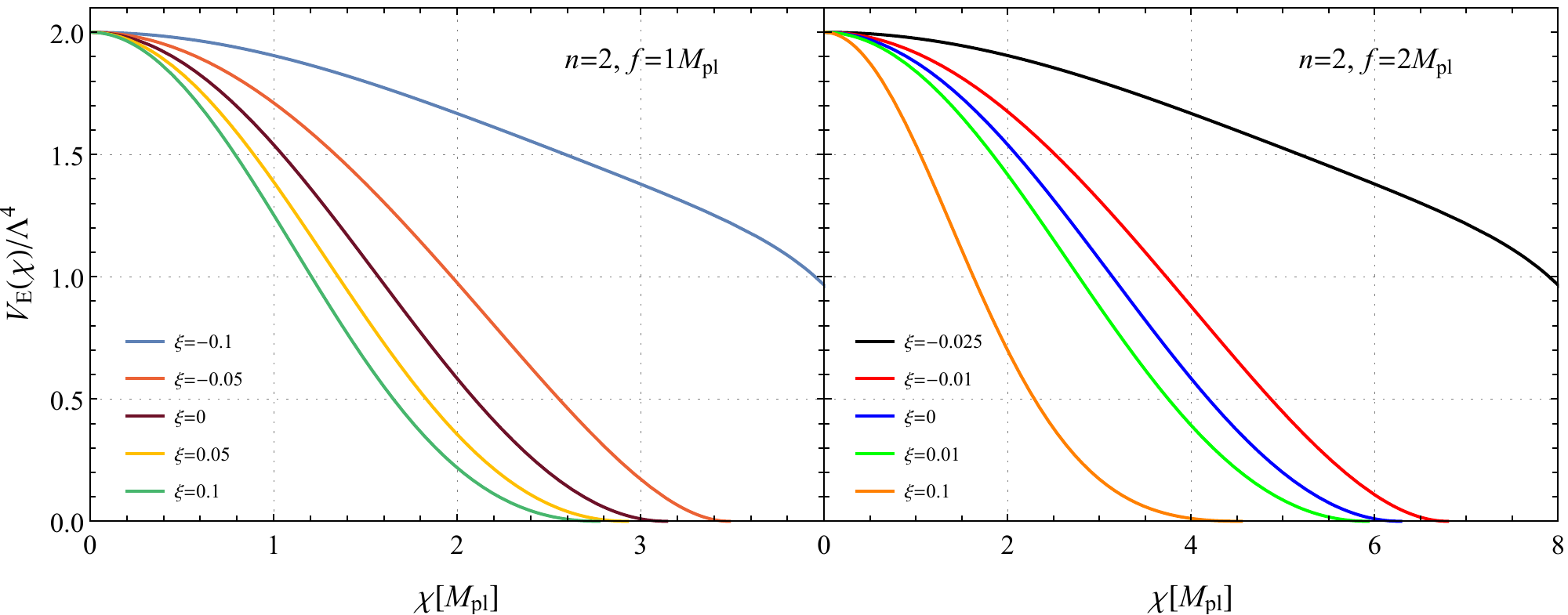}  \\ \vspace{5mm}
      \includegraphics[width= 0.9\textwidth]{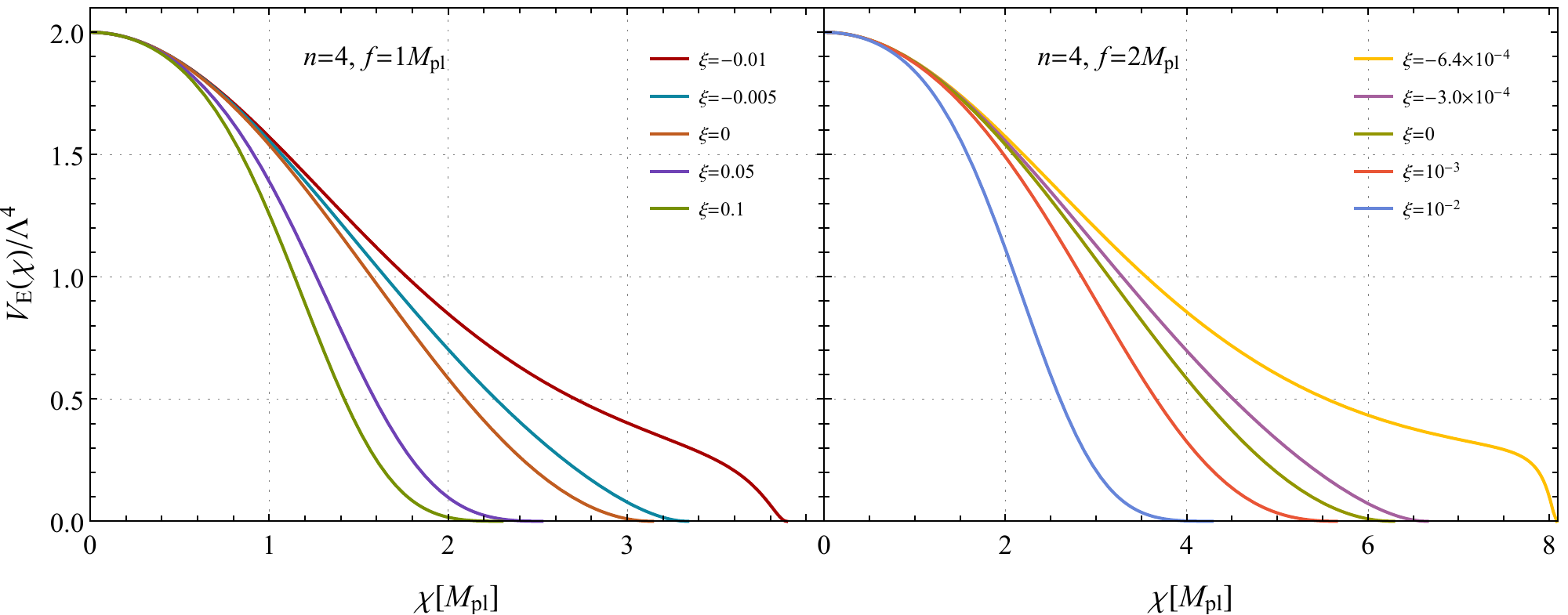}
  \caption{
    Ratio of the Einstein frame potential to fourth power of its scale, $V_\mathrm{E}/\Lambda^4$, as a function of $\chi$ in the Palatini formulation of gravity. The changes with different values of the coupling constant $\xi$ are indicated by different color lines, and changes with respect to different values of $n$ (or $f$) can be seen by comparing the curves in top-bottom (or left-right) frames.}
  \label{fig:VechiPal}
\end{figure}

The relations between $\chi$ and $\phi$ are depicted in \fig{chiphipal}.
It again confirms that $\chi$ is a strictly increasing function of $\phi$ and that the response of the curves for $n=4$ is more sensitive than that of the case $n=2$. Compared to the metric theory, the increase of the potential is mild due to the reason that the second term in \Eq{conftran} does not contribute in this case. The same reason results in less flat potential in the Palatini theory than that in the metric theory. This property can also be clearly seen in \fig{VechiPal} by comparing it with \fig{Vechi}. 

\begin{figure}[h]
  \centering
  \includegraphics[width=0.48\textwidth]{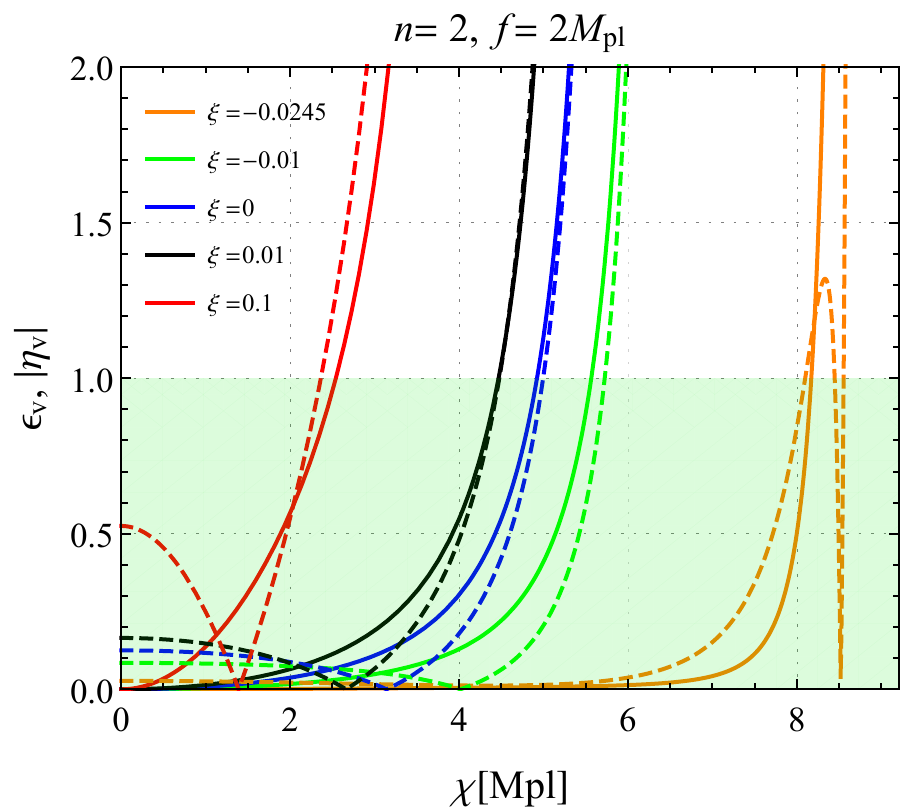} \hspace{2mm}
    \includegraphics[width=0.48\textwidth]{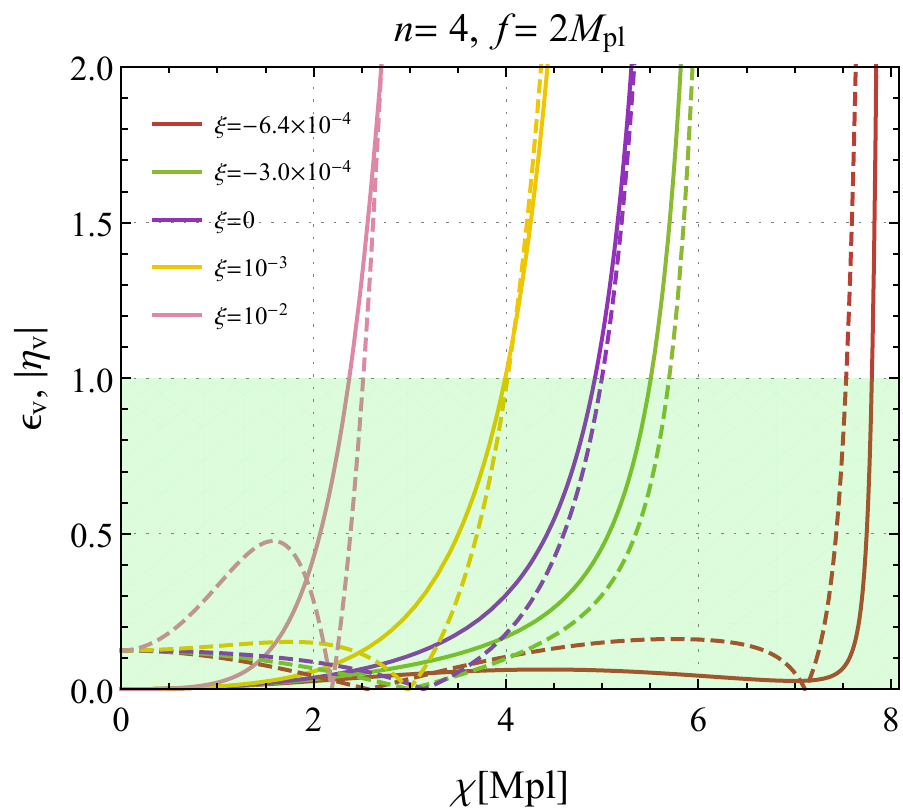}
  \caption{
    The slow-roll parameters $\epsilon_\mathrm{v}$ (solid lines) and $\eta_\mathrm{v}$ (dashed lines) as functions of the inflaton $\chi$ for the fixed values of $n=2,4$ and $f=2M_{\mathrm{pl}}$. When a curve crosses the upper edge of the shaded band, slow-roll conditions are violated and inflation ends.}
  \label{fig:slrfigsPal}
\end{figure}

The slow-roll parameters in this case are
\begin{equation}
 \begin{aligned}
  \epsilon_\mathrm{v} = & \frac{\left[ (F-1)\left(2nV/\Lambda^4 + \frac{x}{\hat{f}} \sin \frac{x}{\hat{f}}\right) + \frac{x}{\hat{f}} \sin \frac{x}{\hat{f}} \right]^2 }{2 x^2 (V/ \Lambda^{4})^2 F } \\ 
                      = & \frac{\left[2  n \xi  x^n+ \left(1+\xi  x^n \right) \frac{x}{\hat{f}} \sin \left(\frac{x}{\hat{f}}\right) +2  n \xi  x^n \cos \left(\frac{x}{\hat{f}}\right)\right]^2}{2 x^2 \left( 1 + \xi  x^{n}  \right)\left[1+ \cos \left(\frac{x}{\hat{f}}\right) \right]^2},
  \end{aligned}
\end{equation}
and 
\begin{equation}
 \begin{aligned}
  \eta_\mathrm{v} = & \frac{ B }{2 x^2(V/ \Lambda^{4}) F},
\end{aligned}
\end{equation}
where
\begin{equation}
\begin{aligned}
 B= & \ n (F-1) \left[4 (1-n)+7\frac{x }{\hat{f} }\sin \frac{x }{\hat{f} }+(F -1) \left(4+6 n+7\frac{ x }{\hat{f} } \sin \frac{x }{\hat{f} }\right)\right] \\
 &-2 \cos \frac{x }{\hat{f} }\left[\frac{x^2 F^2}{\hat{f}^2}+n (F -1) [2 (n-1)-(2+3 n) (F-1)]\right].
 \end{aligned}
\end{equation}
As before, we omit variables in $F(\xi,\phi)$ and potential $V(\phi)$. The slow-roll parameters do not depend on the scale $\Lambda$. Field space satisfying the slow-roll conditions are depicted in \fig{slrfigsPal}. As is shown in this figure, comparing to the metric case, field space gets shrunk considerably, which means the allowed range that the inflaton $\chi$ can move is small, and one may worry about there is no enough field space for sufficient e-folds when $n=2$ in this Palatini approach, let alone $n=4$. However, we show later that there is still enough field space to do inflation. Both left and right plots indicate that when $\xi$ stays very close to zero, say $\left| \xi \right| < 10^{-2}$ for $n=2$ and $\left| \xi \right| < 10^{-3}$ for $n=4$, $\epsilon_\mathrm{v}$  violates the slow-roll conditions first, while for a larger deviation $\eta_\mathrm{v}$ violates first.

\begin{figure}[h]
  \centering
  \includegraphics[width=0.49\textwidth]{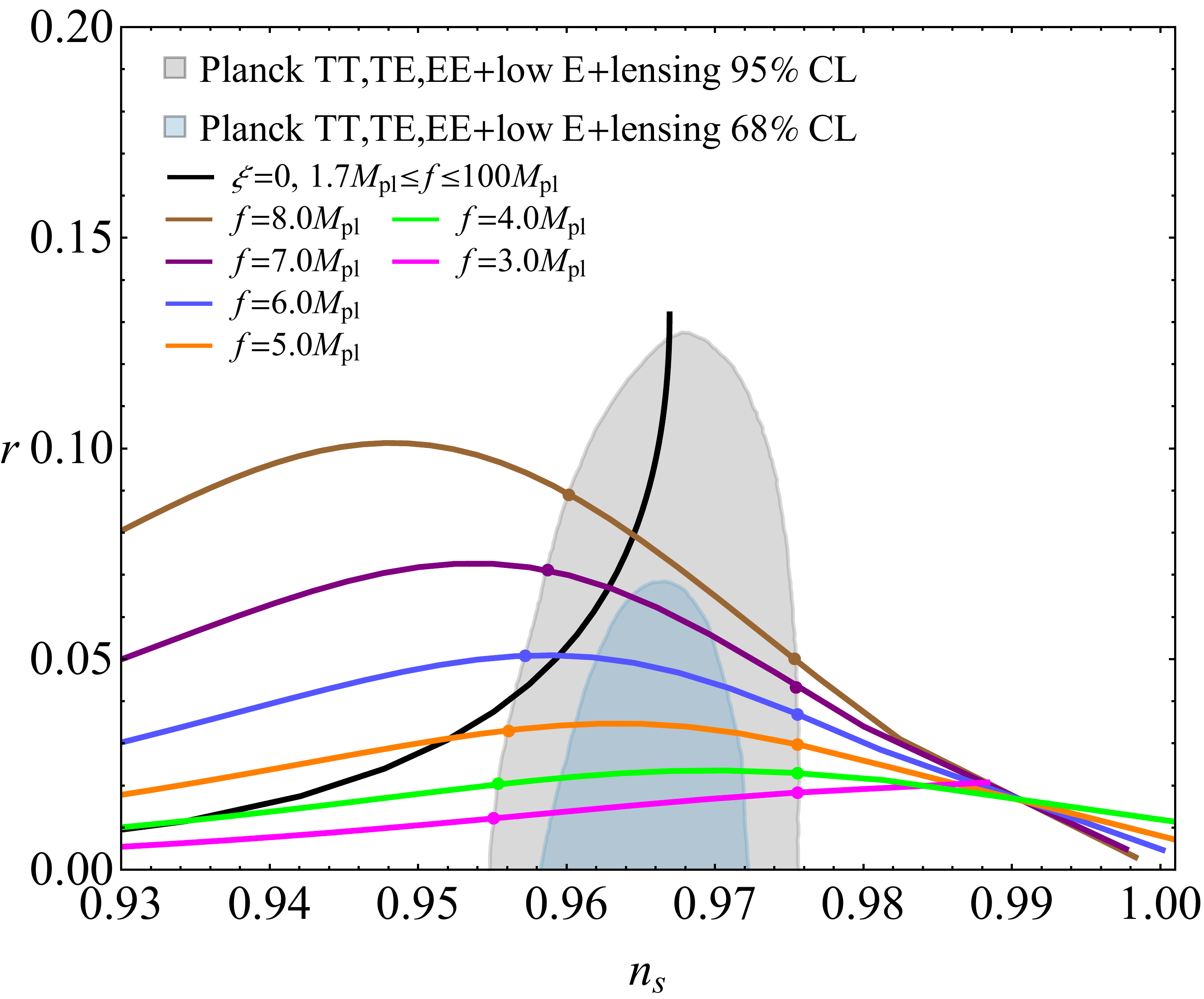}\hspace{1.5mm}
  \includegraphics[width=0.49\textwidth]{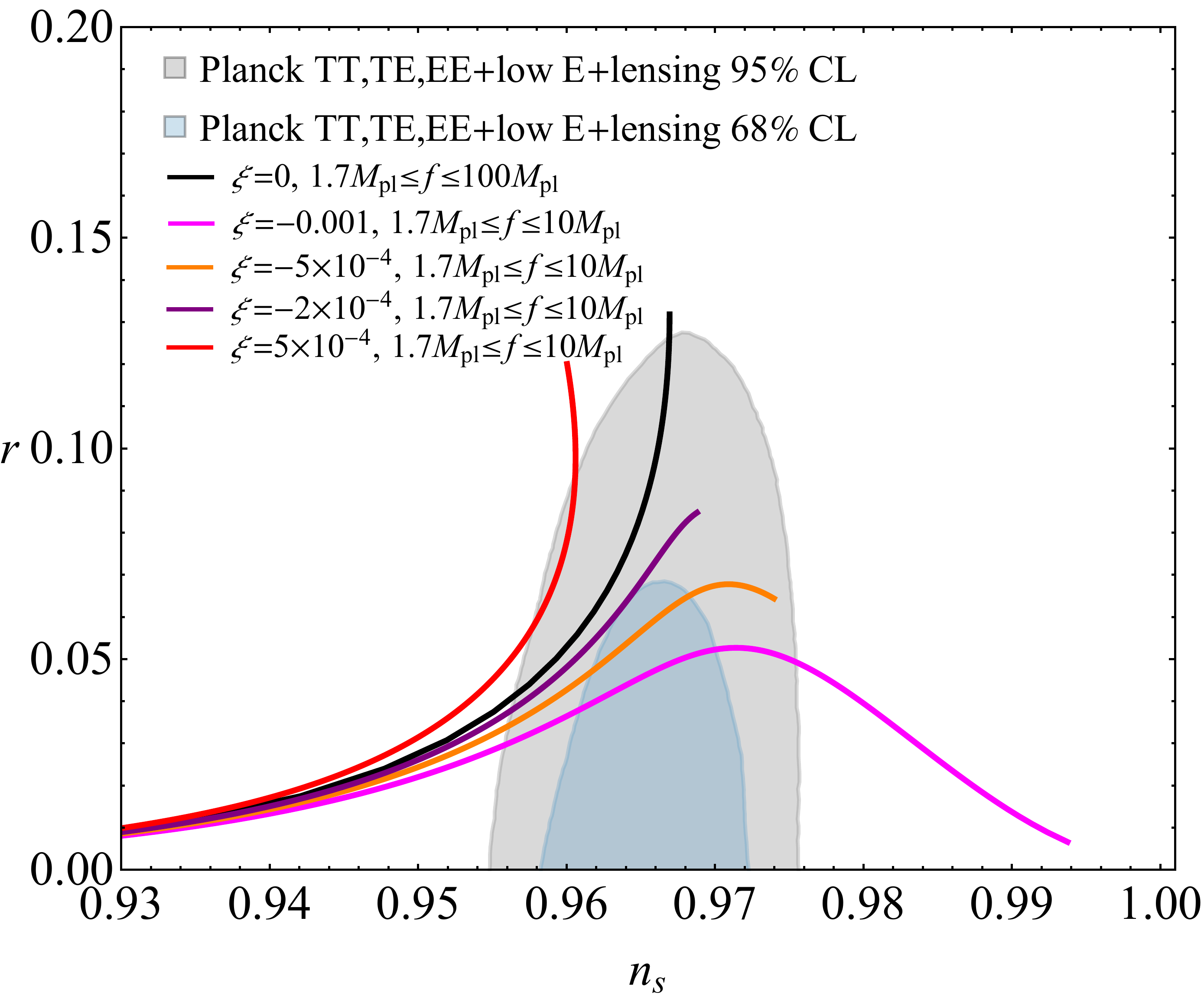}
  \caption{Plots of $r$ vs  $n_s$ obtained by the natural inflation with a nonminimal coupling to gravity in the Palatini theory. This figure shows the case where $n=2$ and 60 e-folds are achieved during the inflation.}
  \label{fig:nsrplotPal}
\end{figure}

Predictions on the cosmological observables $n_s$ and $r$ in the Palatini theory are shown in \fig{nsrplotPal}, which are similar to the corresponding results in the metric theory. The reason is the following. Since a very weak coupling (for instance,  $|\xi|\sim \mathcal{O}(10^{-3})$ when $n=2$) is needed for the rehabilitation of the natural inflation in our discussions, the first term of the expression $\frac{d\chi}{d\phi}$ in \Eq{conftran} has a dominant contribution to the derivative, and the second term gives a mild correction in the entire field space. The latter, being the only discriminator for the inflations between the metric and the Palatini theories, does not give rise to a significant change in $\chi_{*}$. Thus predictions on $n_s$ and $r$ are hardly distinguishable in these two theories.

A rather clear distinction between the results of the two theories is the following. The lowest energy scale for $f$ in the Palatini theory is $1.95M_{\mathrm{pl}}$, above which the values of $n_s$ and $r$ stay within $95\%$ C.L. region of the Planck data. While in the case of metric theory, its value is a bit smaller, $1.9M_{\mathrm{pl}}$. Since the results are similar to the metric theory and $n>2$ case deteriorates the agreement between the predictions and data, results from the higher powers of $\phi$ in the coupling function $F(\xi, \phi)$ are not illustrated any further.

\section{Discussion of the results and reheating after the inflation}
\label{sec:discussions}

The previous sections show that negative couplings bring the curves of $r$ with respect to $n_s$ toward the preferred region of the Planck CMB measurement, whereas positive couplings push them to the excluded areas. We get a lower bound of $f\geqslant1.9M_{\mathrm{pl}}$ in the metric case and $f\geqslant 1.95M_{\mathrm{pl}}$ in the Palatini case from the intersection of the $(n_s, r)$ curve with the $95\%$ CL contour of the Planck data. We also find an upper bound for $f$ around $30 M_{\mathrm{pl}}$. Now, we answer a related question: can we obtain a range of the coupling $\xi$, for a given $f$, which makes predictions for $(n_s, r)$ inside the $95\%$ CL region? Such a range can be found from \fig{nsrplot} and \ref{fig:nsrplotPal} by identifying the intersection points of the curves and the Planck $95\%$ contours. Results are summarized in \tab{paramcollect}. Allowed ranges of $\xi$ are rather narrow. They are all of the order of $10^{-3}$, and the bound on $\xi$ gets tighter with the increase of $f$.
\begin{table}
\centering
\renewcommand{\arraystretch}{1.8}
\begin{tabular}{|c|c|c|}
  \hline
  $f (M_{\mathrm{pl}})$ & $\xi (10^{-3})$ metric case &  $\xi (10^{-3})$ Palatini case\\ \hline
   $3$ & $[-10.72, -8.70]$ &  $[-10.78, -8.72]$ \\ \hline
   $4$ & $[-5.52, -3.02]$  &  $[-5.54, -3.01]$  \\ \hline
   $5$ & $[-3.20, 0.60]$   &  $[-3.19, 0.61]$   \\ \hline
   $6$ & $[-2.02, 0.40]$   &  $[-2.03, 0.36]$   \\ \hline
   $7$ & $[-1.41, 0.62]$   &  $[-1.39, 0.61]$   \\ \hline
   $8$ & $[-0.97, 0.55]$   &  $[-0.97, 0.54]$   \\ \hline
\end{tabular}
\caption{Ranges of the coupling $\xi$ for given values of symmetry breaking scale $f$. The bounds come from $95\%$ CL region of the Planck 2018 result. 
}
\label{tab:paramcollect}
\end{table}

\begin{figure}
  \centering
  \includegraphics[width=0.7\textwidth]{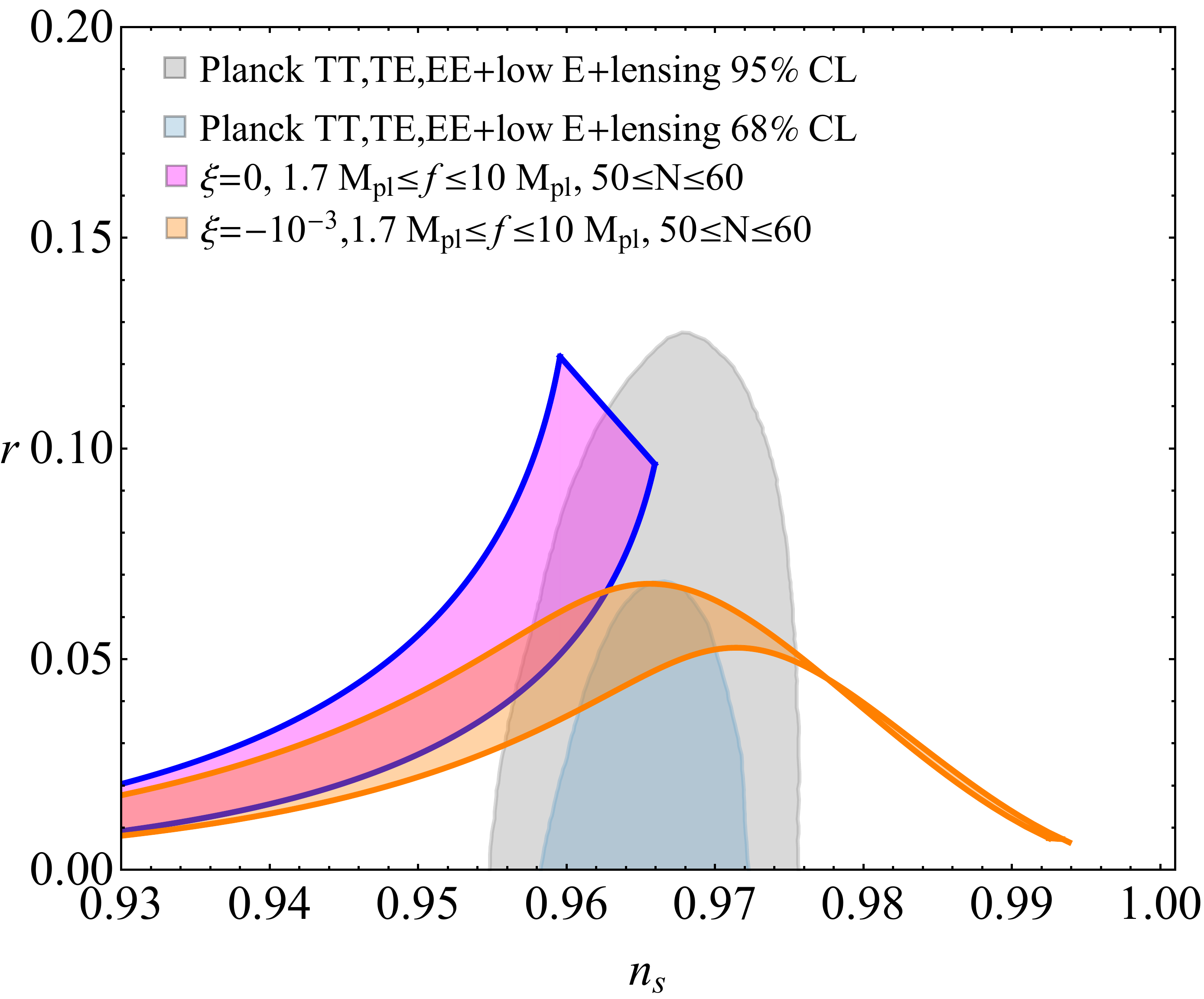}\hspace{1.5mm}
  \caption{Cosmological observables $n_s$ and $r$ obtained by natural inflation with and without the nonminimal coupling. This figure shows the case where $n=2$ and 50-60 e-folds are achieved during the inflation.}
  \label{fig:nsrplotMetPal}
\end{figure}

Previously, all of our results are presented for achieving a rigid 60 e-folds before the end of inflation. Considering the uncertainties in the details of the reheating process and post-inflationary thermal history of the Universe, we also examine the validity and stability of our results in the range of 50-60 e-folds. For illustrative purpose and also for comparison between the minimal and nonminimal coupling scenarios, we fix $\xi = 0$ as well as $\xi = -10^{-3}$ but change $f \in [1.7, 10.0]M_{\mathrm{pl}}$, and show the results in~\fig{nsrplotMetPal}. As we can see, $n_s$ and $r$ have different behaviors in these two scenarios: both $n_s$ and $r$ increase with the increase of $f$ in the minimal coupling case~\footnote{Although this statement is a bit implicit, its meaning can be understood from the $N=50$ and $N=60$ lines, which are upper and lower edges of the area. The left ends of the lines correspond to small $f$ and the right ends correspond to large $f$. These lines are generated from left to right by increasing the value of $f$.}, while in the case of a nonminimal coupling $n_s$ increases faster than that of the minimal coupling case, and $r$ increases in the beginning then decreases. What is more, the predicted area from the minimal coupling has partial overlap with the $95\%$ CL contour but never intersects with the $68\%$ CL contour of the Planck data. In contrast, the area obtained with nonminimal coupling can overlap with both of these contours.

So far the scale $\Lambda$ does not enter the expressions of these observables: it neither affects the number of e-folds nor the slow-roll parameters. But it is constrained by the power spectrum of the curvature perturbation. The power spectrum can be expanded as
\begin{equation}
 P_\mathcal{R}(k) = A_s(k)\left( \frac{k}{k_*} \right)^{n_s-1},
\end{equation}
where $A_s(k)$ is the amplitude and $k_*$ is a pivot scale. The Planck 2018 TT, TE, EE+ LowE+lensing data  provides $\ln \left(10^{10}A_s\right)= 3.044\pm 0.014$~\cite{Akrami:2018odb}. The power spectrum, on the other hand, is computed at the horizon crossing scale,
\begin{equation}
 P_\mathcal{R} = \left( \frac{H^2}{2\pi \dot{\chi}}\right)^2 \simeq \frac{V_\mathrm{E} }{24 \pi^{2} M_{\mathrm{pl}}^{4} \epsilon_{\mathrm{v}}}.
\end{equation}
Approximating this quantity with the Planck result, we extract the constraints on $\Lambda$, which are shown in \tab{lambdamass}. To get these numbers, we use the ranges of $\xi$ listed in \tab{paramcollect}. From these results, we notice that the scale $\Lambda$ is constrained to be around $10^{-3}M_{\mathrm{pl}}$, which is below the grand unification scale, $10^{16}$ GeV.
\begin{table}
\centering
\renewcommand{\arraystretch}{1.8}
\begin{tabular}{|c|c|c|}
  \hline
  $f (M_{\mathrm{pl}})$ & $\Lambda$ ($10^{-3} M_{\mathrm{pl}})$&  $m_\chi$ ($10^{13}$ GeV) \\ \hline
   $3$ & $[3.75, 4.23]$ &  $[1.14, 1.45]$ \\ \hline
   $4$ & $[4.28, 4.52]$  &  $[1.11, 1.24]$  \\ \hline
   $5$ & $[4.70, 5.02]$   &  $[1.07, 1.22]$   \\ \hline
   $6$ & $[5.28, 5.59]$   &  $[1.13, 1.27]$   \\ \hline
   $7$ & $[5.60, 6.23]$   &  $[1.09, 1.35]$   \\ \hline
   $8$ & $[6.00, 6.85]$   &  $[1.09, 1.42]$   \\ \hline
\end{tabular}
\caption{Ranges of the scale $\Lambda$ as well as the mass of the field $\chi$ for given values of $f$ and ranges of $\xi$ in \tab{paramcollect}. As the Palatini case gives very similar results, here we report the results from the metric case.
}
\label{tab:lambdamass}
\end{table}
The ranges of $\Lambda$, in turn, constrain zero temperature mass of the inflaton, $m_{\chi} = \Lambda^2/f$, to be around $10^{13}$ GeV (see last column of the \tab{lambdamass} for its ranges). 

We show that natural inflation with a nonminimal coupling to gravity is still a candidate of viable models because it ends gracefully, through violation of the slow-roll conditions, and its predictions are consistent with data. To solidify this argument, we consider the last important property, which is a successful reheating. In the usual reheating mechanism, after the end of inflation, the inflaton field oscillates around the minimum of the potential, and it can decay into light particles such that its energy density is converted into that of radiation. Through this way, the Universe gets reheated. As the potential in our model possesses a minimum, the inflaton can oscillate around the minimum and lead to a usual reheating. On top of this, reheating can also be realized by gravitational particle production~\cite{Ford:1986sy} when the kinetic energy $\frac{1}{2}\dot{\chi}^2$ of the inflaton dominates in the energy density $\rho = \frac{1}{2}\dot{\chi}^2 + V_\mathrm{E}(\chi)$. The domination of the kinetic energy can be quantitatively studied with the equation of state parameter $\omega$ of the inflaton field, which is
\begin{equation}
 \omega \equiv \frac{p}{\rho}=\frac{\frac{1}{2}\dot{\chi}^2 - V_\mathrm{E}(\chi)}{\frac{1}{2}\dot{\chi}^2 + V_\mathrm{E}(\chi)},
\end{equation}
where the pressure $p = \frac{1}{2}\dot{\chi}^2 - V_\mathrm{E}(\chi)$. One can see that $\omega \approx -1$ if the kinetic energy is negligible and $\omega \approx 1$ if it is the dominant component of the energy. On the other hand, the equation of state parameter $\omega$ controls the evolution of the energy density
\begin{equation}
 \rho \propto a^{-3(1+\omega)},
\end{equation}
where  $a$ is the scale factor. It is pointed out that a change in the spacetime metric at the end of inflation creates particles due to their coupling to the spacetime curvature, this type of particle production is called gravitational particle production~\cite{Parker:1969au,Ford:1986sy}. The gravitational particle production mechanism works when the kinetic energy of the inflaton dominates over the potential energy, which is called kination. In the kination epoch, $\omega \simeq 1$, the energy density of the inflaton field $\rho \propto a^{-6}$, which drops faster than that of the radiation, $\rho_\mathrm{rad} \propto a^{-4}$, as the Universe expands. As a result, the energy density of the radiation produced by the gravitational particle production dominates, and the Universe enters the radiation dominated era. Now we show that this mechanism is applicable in our model.

\begin{figure}
  \centering
  \includegraphics[width=0.6\textwidth]{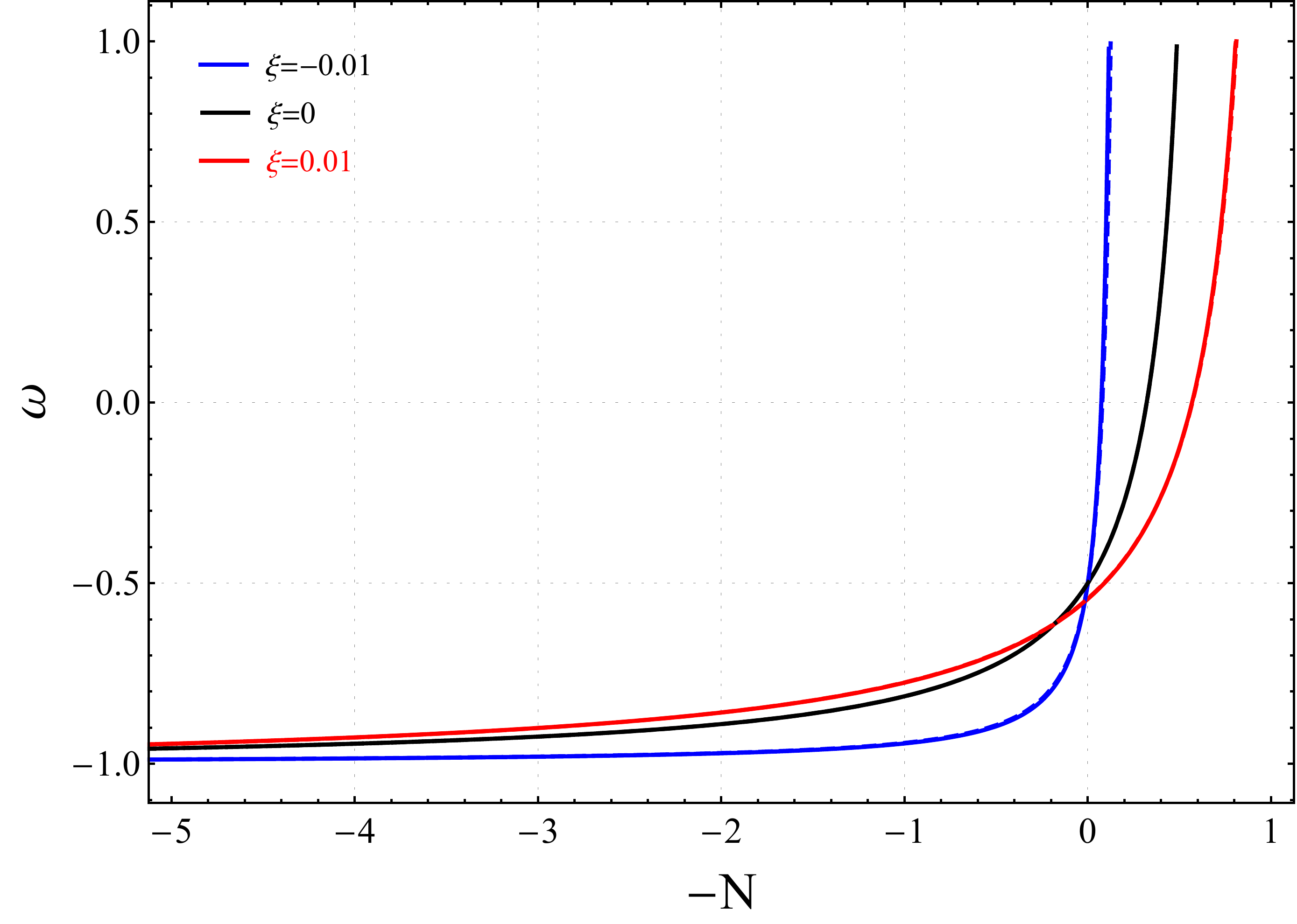}\hspace{1.5mm}
  \caption{Behavior of the equation of state parameter $\omega$ as a function of the number of e-folds, using  three different values of the coupling $\xi$. On the horizontal axis, the 0 corresponds to the end of inflation, negative values correspond to the inflationary epoch, and the positive values indicate post-inflationary epoch. In the plot we use $n=2$, $f=3M_{\mathrm{pl}}$, in both the metric (solid line) and the Palatini (dashed lines) cases, their differences are very small. The results from other values of $n$ and $f$ are similar.}
  \label{fig:eqstparMetPal}
\end{figure}
In \fig{eqstparMetPal} we show the evolution of $\omega$ as a function of the number of e-folds, counting from the end of inflation, with different benchmark values of the coupling $\xi$. The zero point in the horizontal axis corresponds to when the inflation ends, negative numbers of e-folds indicate the inflationary stage, and positive values indicate the post-inflationary stage. We use $n=2$, $f=3M_{\mathrm{pl}}$ for illustration, and check that for other values of these parameters the qualitative behavior of $\omega$ does not change much. Results from the metric approach (with solid lines) resemble that from the Palatini approach (with dashed lines) due to the reasons mentioned earlier. As we can see, $\omega$ is very close to -1 during inflation which means the kinetic energy of the inflaton field  is negligible, its energy is dominated by the potential energy. The kinetic energy  increases very quickly after the end of inflation, and $\omega$ gets very close to its maximum value within one e-fold. Therefore, in our model, the gravitational reheating is an efficient way to convert the inflaton energy to the radiation bath.  

\section{Conclusions}
\label{sec:conclusions}

We have discussed one of the simplest extensions of natural inflation, where the inflaton field has a nonminimal coupling to gravity in both frameworks of the metric and the Palatini theories. We start with a general form of the coupling function and then extensively study its consequences in a weak coupling regime. For simplicity,  we choose to use $F(\xi, \phi) = 1+\xi\left( \frac{\phi}{M_{\mathrm{pl}}} \right)^n$ with $n=2,4,\dots$, and carry out our analysis in the Einstein frame. The lowest power in this function gives the best agreement with allowed regions in $(n_s, r)$ plane from the Planck temperature, polarization, and lensing data. Confronting the model predictions with the Planck 2018 results, we put constraints on the model parameters. It is worth mentioning that the parameter space of the model is testable in the future observations, such as BICEP3~\cite{Wu:2016hul} and the Simons Observatory~\cite{Ade:2018sbj} which are expecting to set upper limits on $r$ in the percent and permill levels, respectively.

We find that the symmetry breaking scale $f$ has to be larger than $1.9 M_{\mathrm{pl}}$ ($1.95 M_{\mathrm{pl}}$) in the metric (Palatini) approach in order to be consistent with the Planck 2018 result. For given values of $f$, the coupling $\xi$ is confined to a small range around $10^{-3}$ to make the predictions stay inside the Planck-allowed region. The model yields predictions on $n_s$ and $r$ inside the $95\%$ CL contour (and also inside the $68\%$ CL contour in the part of parameter space) of the Planck result. Furthermore, since the inflaton couples weakly to gravity, the metric and the Palatini approaches provide very similar results. In this weak coupling regime, a negative coupling flattens the potential and brings predictions to a good agreement with data, whereas a positive coupling makes the situation worse.

As is shown in our discussion, the model accommodates all three important ingredients of a
successful model: a graceful exit is realized by the evolution of the inflaton field, predictions on $n_s$ and $r$ are consistent with the data, and successful reheating is also
possible. Our findings show that although the minimally coupled natural inflation is disfavored by the Planck 2018 data, its slight extension with a nonminimal coupling is still a good candidate of successful inflation models. Apart from this, there are also studies about another possible modification to natural inflation, by introducing an inflaton coupling to a thermal bath \cite{Mohanty:2008ab,Visinelli:2011jy,Mishra:2011vh,Reyimuaji:2020bkm}. They conclude that the thermal effect can bring natural inflation's predictions to a good agreement with data. Although original natural inflation suffers from a strong tension with observations, its variants are still in agreement with the Planck CMB data. Therefore, it is important to test these variants before completely excluding the natural inflation.

Finally, minimally coupled natural inflation agrees with CMB measurements only when $f\gtrsim 5M_{\mathrm{pl}}$. In this paper, we show that introducing a nonminimal coupling can lower this bound to be as small as $2M_{\mathrm{pl}}$, but no further reduction is achieved. This is because a small negative coupling in need has a lower limit $\xi > -\left( \frac{M_{\mathrm{pl}}}{f\pi}  \right)^n$. With smaller values of $f$ and $\xi$ in its ranges, the curves of $r$ with respect to $n_s$ cannot enter the Planck allowed regions.  The super-Planckian $f$ is a drawback of the original natural inflation model and the model in this work. The inflaton potential may get a large correction from the quantum gravity effects for super-Planckian $f$; thus it becomes difficult to realize the model in a more fundamental theory. A work for reducing $f$ to sub-Planckian scales and a UV complete natural inflation model construction will be carried out in the future.

\section*{Acknowledgements}
Y. R. is grateful to the support from the postdoctoral research fellowship of China, and X. Z. is supported by China Postdoctoral Science Foundation under Grant No. 2019M650001.




\end{document}